\documentclass[16pt, usenatbib]{mn2e}
\usepackage{graphicx,times,subfigure}
\usepackage{bm}
\usepackage{epsf}
\usepackage{amsmath}
\usepackage{amssymb}
\usepackage{cases}
\usepackage{float}
\usepackage[caption = false]{subfig}
\usepackage{colortbl}

\usepackage{multirow}

\def\la{\langle}
\def\ra{\rangle}

\def\be{\begin{equation}}
\def\ee{\end{equation}}
\def\ben{\begin{eqnarray}}
\def\een{\end{eqnarray}}

\def\oh{\hat\Omega}
\def\myC{{\cal C}}

\def\bk{{\bf k}}

\def\bk{\bf k}

\def\fsky{ f_{\rm sky}}
\def\fNL{ f_{\rm NL}}
\def\fNLhat{ \hat{f}_{\rm NL}}

\newcommand*{\llangle}{\langle\kern-2\nulldelimiterspace \langle}
\newcommand*{\rrangle}{\rangle\kern-2\nulldelimiterspace \rangle}
\begin{document}
\onecolumn
\title[Principal Components of CMB non-Gaussianity]
{Principal Components of CMB non-Gaussianity}
\author[Regan \& Munshi]
{Donough Regan \& Dipak Munshi\\
Astronomy Centre, School of Mathematical and Physical Sciences, University of Sussex, Brighton BN1 9QH, United Kingdom\\}
\maketitle
\begin{abstract}
The skew-spectrum  statistic introduced by \cite{MuHe10} 
has recently been used in studies of non-Gaussianity from diverse cosmological data sets
including the detection of primary and secondary non-Gaussianity of Cosmic Microwave Background (CMB) radiation.
Extending previous work, focussed on independent estimation, here we deal with the question of joint estimation of 
multiple skew-spectra from the same or correlated data sets. 
We consider the optimum skew-spectra for various models of primordial non-Gaussianity as well as
secondary bispectra that originate from
the cross-correlation of secondaries and lensing of CMB: coupling of lensing with the Integrated Sachs-Wolfe (ISW) effect,
coupling of lensing with thermal Sunyaev-Zeldovich (tSZ), as well as from unresolved point-sources (PS). 
For joint estimation of various types of non-Gaussianity, we use the PCA to construct the linear combinations of 
amplitudes of various models of non-Gaussianity, e.g. $f^{\rm loc}_{\rm NL},f^{\rm eq}_{\rm NL},f^{\rm ortho}_{\rm NL}$
that can be estimated from CMB maps. {We describe how the bias induced in the estimation of primordial non-Gaussianity due to
secondary non-Gaussianity may be evaluated for arbitrary primordial models using a PCA analysis. }The PCA approach allows one to infer approximate (but generally accurate) constraints using CMB data sets on any reasonably smooth model by use of a lookup table and performing a simple computation. This principle is validated by computing constraints on the DBI bispectrum using a PCA analysis of the standard templates.
\end{abstract}
\section{Introduction}
The cosmic microwave background (CMB) radiation is the most important probe 
of the very earliest stages of the Universe. In standard inflationary models,
the early Universe should be very close to random Gaussian field. 
Deviations from pure Gaussian statistics can provide direct clues regarding
inflationary dynamics \citep{Bartolo04,Komat10,Lig10}.

Analysis of temperature maps from the nominal Planck mission\footnote{http://sci.esa.int/planck/} (\cite{Planck13}) has set unprecedented
constraints on various models of primordial non-Gaussianity by improving earlier results
from WMAP\footnote{http://map.gsfc.nasa.gov/} (\cite{WMAP9}). The Planck team also reported detections of secondary non-Gaussianity
generated by coupling of the Integrated Sachs Wolfe (ISW) effect with lensing of the CMB, as well as
from residual unresolved point sources. Following the data release from the Planck team 
cross-correlation of thermal Sunyaev-Zeldovich (tSZ) effect and lensing of CMB has also been reported by \cite{HS14}.

Study of non-Gaussianity is typically performed using multiple techniques for cross validation. 
The techniques include the optimal KSW (\cite{KSW}) estimator, skew-spectrum (\cite{MuHe10}), modal decomposition (see \cite{Modal10} and references therein)
as well as the sub-optimal Minkowski functional (see \cite{Ducont13} and references therein). The skew-spectrum was designed to address the dual challenge of estimation of {\it primary} ({produced in the early universe during the inflationary epoch}) or {\it secondary} ({arising from late-time effects after recombination, e.g. due to lensing}) non-Gaussianity
without compressing all the available information into a single number in an optimal way.  {More precisely, the skew-spectrum involves a data compression of the $\ell_{\rm max}^2$ independent modes of the bispectrum down to $\ell_{\rm max}$ numbers (where $\ell_{\rm max}$ is determined by the resolution of the experiment under consideration, and is $\mathcal{O}(1500-2000)$ for the Planck satellite).} It has the potential to differentiate
among various possible sources of non-Gaussianity as well as contamination from unknown systematics. Initially
proposed for the detection of non-Gaussianity via the bispectrum, the method has now been extended to include non-Gaussianity 
at the level of tripsectrum in \cite{MuHeCoSmCoSe11}. The skew-spectrum, being a (pseudo) power-spectrum associated with the underlying
bispectrum, can provide an estimate of non-Gaussianity at each harmonic mode $\ell$. 
However, not only are the individual skew-spectrum modes generally correlated for a given underlying model of bispectrum,
different skew-spectra estimated from the same data may be correlated too. 
Based on a Fisher matrix analysis, in this paper we are primarily interested in finding independent linear combinations
of different modes of the same skew-spectrum as well as among harmonics associated
with different skew-spectra. In doing so we will adopt the well established technique of 
principal components analysis (PCA) in our study.

In recent years there has been
a renewed interest in applying principal component analysis techniques to
various cosmological data sets, a technique pioneered by \cite{EB99}.
Extending previous work here we consider the joint estimation of different types of
non-Gaussianity from the same data set. 
This method can reveal the detailed statistical structure of
parameter space which is lacking in an one-dimensional confidence
level presentation. {For example in \cite{Rocha04}
the possibility of measurement of the fine-structure constant $\alpha$ was explored in the context of CMB data with analysis based on Fisher matrix and PCA. 
Principal component analysis was applied to the decorrelation of the power spectrum of galaxies in \cite{HT00}, while, more recently, the technique has been applied to
study of reconstruction of reionization history \cite{MH08} as well as inflationary potential reconstruction \cite{DH10}. In the context of primordial non-Gaussianity, PCA was applied to a modal decomposition of an effective field theory description of single-field inflation in \cite{ARS14}, demonstrating that only four linearly independent combinations may be constrained using WMAP data. The approach allows a certain independence from the primordial templates used for comparison to the data, with the principal directions corresponding to those shapes that may be best constrained. 
}

{In this work, the PCA approach is utilised in the context of the skew-spectrum, and extended to include secondary shapes. We will perform $two$ separate principal component analyses. The first will concern the decomposition of the skew-spectrum associated with individual bispectra - both primordial and secondary. In so doing, we wish to uncover the number of independent modes present in the skew-spectrum and whether a more dramatic data compression may be viable. (Note that for a maximum constrainable harmonic mode, $\ell_{\rm max}$, the bispectrum has $\mathcal{O}(\ell_{\rm max}^2)$ modes, while the skew-spectrum has $\mathcal{O}(\ell_{\rm max})$ modes.). }

{The second PCA decomposition will concern the joint estimation of various bispectra. This approach allows us to identify orthogonal directions and includes the analysis of several primordial and secondary bispectra. Given an arbitrary (smooth) model, $M$, this procedure allows for an efficient approach to obtain its constraints from a given data set without having to reproduce the entire analysis pipeline. In particular, given the constraints on standard templates, one need only carry out the Fisher analysis of these templates (and the model itself), identify the principal components, $P_i$, and correlate these with that of the model. In this fashion one may utilise the constraints on the standard template, to infer the constraints on the model of interest. This approach is easily adopted, in the case of a primordial model, to the calculation of the bias induced by any secondary model\footnote{{For clarity, the bias refers to a non-zero estimation for the bispectrum of a primordial model due solely to the overlap of the shape of that model with a secondary model. It is important to subtract this bias in order to correctly estimate the amplitude of the primordial bispectrum.} }.
The accuracy is dependent on a dense enough sampling of the parameter space. However given the analysis of \cite{ARS14} identifying only four principal directions in the case of single-field inflation, the density is clearly rather low in general for smooth bispectra (i.e. bispectra without sharp features). As a concrete example of this approach we compute constraints on the DBI model, as well as the bias induced by three different secondary sources of non-Gaussianity. This is achieved using the principal components amongst the three standard primordial templates (local, equilateral, orthogonal). We further, present a PCA analysis of the skew-spectrum associated with each principal direction. While these principal directions are orthogonal to one another, each may themselves contain important harmonic mode information for a more exacting identification of the angular scales at which the various skew-spectra have their highest signal-to-noise ratios. 
}

This paper is organised as follows: In \textsection\ref{sec:bispec} we present a brief
review of various models of non-Gaussianity primary and secondary.
We introduce our estimators in \textsection\ref{sec:skew}. After discussing the process involved in performing a joint measurement
of these estimators in \textsection\ref{sec:joints},
we introduce the principal components in \textsection\ref{sec:principal}. We apply our techniques to Planck-like data in \textsection\ref{sec:results}.
\textsection\ref{sec:conclu} is devoted to concluding remarks. 

{{\bf{Notation}:} Symbols denoted using an overhat (e.g. $\hat{B}$) will represent measurements made using the data, or a single realisation of a simulation of the data. The same symbol written without an overhat shall denote the expectation value for the particular model under investigation, or equivalently the average over several realisations of simulations of that model (e.g. $B=\langle \hat{B}\rangle$).}

\section{Primary and Secondary CMB non-Gaussianity}
\label{sec:bispec}

The harmonic transform $a_{\ell m}$ of the temperature fluctuation $\delta {\rm T}(\oh)$ is defined on the surface of the sky as a function of
the angular coordinate $\oh \equiv (\theta,\phi)$:
\ben
&& a_{\ell m} \equiv \int \; d \oh \; Y_{\ell m}(\oh){\delta {\rm T}(\oh) \over {\rm T}}\,,
\een
{where $Y_{\ell m}$ are the basis of spherical harmonics}. 
The angular bispectrum, $B_{\ell_1\ell_2\ell_3}$, is the three-point correlation function defined in the
harmonic domain:
\ben
&& \la a_{\ell_1m_1}a_{\ell_2m_2}a_{\ell_3m_3}\ra = 
\left ( \begin{array}{ c c c }
    \ell_1 & \ell_2 & \ell_3 \\
     m_1 & m_2 & m_3
  \end{array} \right) B_{\ell_1\ell_2\ell_3}.\\
&& B_{\ell_1\ell_2\ell_3} = {b}_{\ell_1\ell_2\ell_3}h_{\ell_1\ell_2\ell_3}; \quad\quad
h_{\ell_1\ell_2\ell_3} \equiv\sqrt{ (2{\ell_1}+1)(2{\ell_2}+1)(2{\ell_3}+1)\over 4\pi  }\left ( \begin{array}{ c c c }
    \ell_1 & \ell_2 & \ell_3 \\
     0 & 0 & 0
  \end{array} \right). 
\een
This form preserves the the rotational invariance of the three-point correlation function in the harmonic domain.
The quantity in parentheses is the Wigner 3j-symbol, which is non-zero only for triplets ($\ell_1,\ell_2,\ell_3$) which satisfy the triangle rule, including
that the sum $\ell_1+\ell_2+\ell_3$ is even,  ensuring the parity invariance of the bispectrum
(see \cite{MuSmCoReHeCo13} for a discussion regarding odd-parity skew-spectrum). 
The reduced bispectrum $b_{\ell_1\ell_2\ell_3}$ was introduced by \cite{KS01} which will be helpful for separating the
purely geometrical factor from the dependence on underlying physics (see \cite{BCP04} for a more detailed discussion).
{
We shall denote by $\hat{B}_{\ell_1\ell_2\ell_3}$, the angle averaged `bispectrum' measured from the {\it data}, given analogously by
\ben
 a_{\ell_1m_1}a_{\ell_2m_2}a_{\ell_3m_3}-\langle a_{\ell_1m_1}a_{\ell_2m_2}\rangle a_{\ell_3m_3}-\langle a_{\ell_1m_1}a_{\ell_3m_3}\rangle a_{\ell_2m_2}-\langle a_{\ell_2m_2}a_{\ell_3m_3}\rangle a_{\ell_1m_1} \equiv\left ( \begin{array}{ c c c }
    \ell_1 & \ell_2 & \ell_3 \\
     m_1 & m_2 & m_3
  \end{array} \right) \hat{B}_{\ell_1\ell_2\ell_3}\,,
\een
where we subtract the terms on the left hand side to account for anisotropic noise and masking (for more details see \cite{KSW}; these terms have zero expectation value since the monopole is subtracted out from the temperature map), with the expectation values taken over Gaussian simulations. The effect of the beam, ${\cal B}_{\ell}$ (see equation (\ref{eq:beam}) for analytical expression) and noise $n_{\ell m}$ may be incorporated in these simulations by the transformation\footnote{Due to an unfortunate historical choice, the symbol $b$ is often used for both the beam and the reduced bispectrum, and therefore, we choose the alternative notation for the beam, ${\cal B}_{\ell}$. It should be noted, however, that the effect of the beam will only be present in an overall rescaling of the power spectrum in our estimator.}  ${a}_{\ell m}\rightarrow \tilde{a}_{\ell m}={\cal B}_{\ell} a_{\ell m}+n_{\ell m}$, where $a_{\ell m}$ is the Gaussian realisation for a ideal experiment with the measured angular power spectrum. Assuming that the noise is Gaussian with angular power spectrum, $n_{\ell}$, the total expected power spectrum, and bispectrum are given by
\begin{align}\label{eq:beam_noise}
\myC_{\ell} &\equiv \langle |a_{\ell m}|^2\rangle \rangle  \rightarrow \tilde{\myC}_{\ell} = {\cal B}_{\ell}^2 \myC_{\ell} + n_\ell \nonumber\\
B_{\ell_1 \ell_2 \ell_3}&\rightarrow \tilde{B}_{\ell_1 \ell_2 \ell_3}={\cal B}_{\ell_1}{\cal B}_{\ell_2}{\cal B}_{\ell_3} B_{\ell_1 \ell_2 \ell_3}
\end{align}
In addition it will be necessary to account for masking of the sky through the inclusion in our estimators of a parameter $\fsky\leq 1$ parametrising the fraction of sky coverage (see \textsection\ref{sec:skew}).
While one may use crude measurements of the skewness to estimate the amplitude of the data bispectrum, in practice the signal to noise from such estimators is too low. Therefore, standard estimation techniques attempt to compare template bispectra to that of the data, and find the corresponding amplitude. We shall describe one such estimator, dubbed the skew spectrum, in \textsection\ref{sec:skew}. For clarity of notation, the bispectrum associated with template model, X, will be denoted $B^{(X)}_{\ell_1\ell_2\ell_3}$. In this section we shall first describe various models of primordial non-Gaussianity, and their associated bispectra, and then proceed to describe a selection of secondary sources of non-Gaussianity. Clearly distinguishing between primary and secondary sources is essential if one wishes to ascribe a inflationary origin to any measured signature of non-Gaussianity. 
}
\subsection{Primary Non-Gaussianity}
The single-field slow-roll model of inflation provides a very small
level of departure from Gaussianity, far below present experimental detection
limits (\cite{Mal03,Acq03}). Many other variants, however, will allow for
a much higher-level of non-Gaussianity (\cite{Komat10}).
Various models of primordial non-Gaussianities are known as {\em local}, {\em equilateral}, {\em orthogonal} or {\em folded} models in the literature.   
Different aspects of the physics of the primordial Universe appear in different shapes of the three- and four-point functions.
\begin{itemize}
\item The ``{\em local}'' model appears in  multi-field models of inflation due to interactions which operate on superhorizon scales (\cite{SB90,Gangui94,Verde00,KS01}). 
\item  The ``{\em equilateral}'' NG, include single-field models with
 non-canonical kinetic term (\cite{chen07,CW09,ACMZ04,SL05,Cheung08,LWW08}), such as k-
 inflation (\cite{ADM99,chen07}) 
 or Dirac-Born-Infield (DBI) inflation (\cite{ST04,AST04})
models characterized by more
 general higher-derivative interactions of the inflaton field,
 such as ghost inflation (\cite{ACMZ04}), and
 models arising from effective field theories (\cite{Cheung08}). 
\item Examples of the class of ``{\em folded}'' (or flattened) NG include: single-field models with non-Bunch-Davies vacuum 
(\cite{chen07,HT08}) and models  with general higher-derivative interactions (\cite{SSZ10,BMR10});
\item ``{\em Orthogonal}'' NG may be generated in single-
 field models of inflation with a non-canonical kinetic
 term (\cite{Ren11,Rib11}), or with general higher-derivative
 interactions. The orthogonal form is constructed in such a way that it is nearly orthogonal
to both local and equilateral forms (\cite{SSZ10,Cheung08,Meer09}). 
\end{itemize}
" The extensions to these models to 
take into account isocurvature modes was considered by \cite{HMHC10}. 
\par
The primordial bispectrum, $B_{\Phi}$ is defined by the three-point function of the gravitational potential, $\Phi$, via 
\ben
&& \langle \Phi({\bk}_1)\Phi({\bk}_2)\Phi({\bk}_3)\rangle = (2\pi)^3 \delta_{\rm 3D}({\bk}_1+{\bk}_2+{\bk}_3)B_{\Phi}(k_1,k_2,k_3)\,;
\een
with the Dirac delta function $\delta_{\rm 3D}$ - arising as a consequence of statistical homogeneity - imposing the triangle condition $\sum_i {\bk}_i = {\bf{0}}$. The harmonic transform of the CMB temperature map, $a_{lm}$ and the primordial gravitational potential, $\Phi$, are related in linear perturbation theory through the correspondence 
\ben
&&a_{\ell m}=4\pi (-i)^{\ell} \int \frac{d^3 k}{(2\pi)^3} \Delta_{\ell}(k) \Phi({\bk})Y_{\ell m}^*(\hat{\bk})\,;
\een
where $\Delta_l(k)$ is known as the transfer function. This relationship may be used to relate the primordial bispectrum with its CMB counterpart. The three standard templates used for CMB analysis are the local, equilateral and orthogonal bispectra, given by:
\ben
&& b^{\rm loc}_{\ell_1\ell_2\ell_3} = 2f^{\rm loc}_{\rm NL}\int r^2 dr \left[ \alpha_{\ell_1}(r)\beta_{\ell_2}(r)\beta_{\ell_3}(r) + {\rm 2 \; cyc.perm}
\right ] \,;\\
&& b^{\rm eq}_{\ell_1\ell_2\ell_3} = -6f^{\rm eq}_{\rm NL}\int r^2 dr \left [\alpha_{\ell_1}(r)\beta_{\ell_2}(r)\beta_{\ell_3}(r) + (2\; \rm cyc.perm.)
- \beta_{\ell_1}(r)\gamma_{\ell_2}(r)\delta_{\ell_3}(r) + (5 {\; \rm cyc. perm.}) + 2 \delta_{\ell_1}(r)\delta_{\ell_2}(r)\delta_{\ell_3}(r)\right ]\,;\\
&& b^{\rm orth}_{\ell_1\ell_2\ell_3} = -6f^{\rm ortho}_{\rm NL}\int r^2 dr \left [\alpha_{\ell_1}(r)\beta_{\ell_2}(r)\beta_{\ell_3}(r) + (2\; \rm cyc.perm.)
- \beta_{\ell_1}(r)\gamma_{\ell_2}(r)\delta_{\ell_3}(r) + (5 {\; \rm cyc. perm.}) + 4 \delta_{\ell_1}(r)\delta_{\ell_2}(r)\delta_{\ell_3}(r)\right ].
\een
The radial functions $\alpha_{\ell}(r)$, $\beta_{\ell}(r)$, $\gamma_{\ell}(r)$ and $\delta_{\ell}(r)$ depend on the power-spectrum of the primordial potential fluctuation  $P_{\Phi}(k)$
and the radiation transfer function $\Delta_{\ell}(k)$:
\ben
&& \alpha_{\ell}(r)  \equiv {2 \over \pi} \int k^2 dk \Delta_{\ell}(k)j_{\ell}(kr); \quad
\beta_{\ell}(r)  \equiv {2 \over \pi} \int k^2 dk \, P_{\Phi}(k)\, \Delta_{\ell}(k)j_{\ell}(kr); \quad\\
&& \gamma_{\ell}(r)  \equiv {2 \over \pi} \int k^2 dk \, P^{1/3}_{\Phi}(k)\, \Delta_{\ell}(k)j_{\ell}(kr); \quad
\delta_{\ell}(r)  \equiv {2 \over \pi} \int k^2 dk \, P^{2/3}_{\Phi}(k)\, \Delta_{\ell}(k)j_{\ell}(kr).
\een
These functions are computed using publicly-available Boltzmann solvers such as CMBFAST \cite{SZ96} or CAMB \cite{LCL00}. {While we shall use the terms {\it model} and {\it template} interchangeably in this paper, we caution that `models' such as the equilateral and orthogonal shapes above are, in fact, templates whose shape has been chosen to provide accurate but separable approximations - i.e. of the form $B_\Phi(k_1,k_2,k_3)\sim X(k_1) Y(k_2) Z(k_3)$ for the one-parameter functions $X,Y,Z$ - to more physically motivated (but non-separable) models. The usage of a principal component analysis over several templates offers the opportunity for more accurate constraints to be placed on possibly non-separable bispectra, without the requirement of reproducing the entire analysis pipeline. This shall be detailed further in Section~\ref{sec:bispecPCA}.}

Any of estimates of the parameters $f^{\rm loc}_{\rm NL}$, $f^{\rm eq}_{\rm NL}$ and $f^{\rm orth}_{\rm NL}$ 
are bound to be correlated. One of the aim 
of this study to investigate using PCA linear combinations of these parameters
that can be estimated with minimum error-bars. 

The current limits from nominal Planck mission are $f^{\rm loc}_{\rm NL} = 2.7 \pm 5.8$, $f^{\rm eq}_{\rm NL} = -42 \pm 75$,
and $f^{\rm ortho}_{\rm NL} = -ˆ'25 \pm 39$ (68 \% CL statistical). These estimates however ignore
the cross-correlation among the $f_{\rm NL}$ parameters (\cite{Planck13}).

These models do not exhaust all options and indeed there are other forms which would probe different 
aspects of the inflationary physics (\cite{CW09,HT08,MX07,Huang08,MossGra07}). A feature of the standard templates is that they may be expressed in a separable form, i.e. in the form $f(k_1)g(k_2)h(k_3)$, which allows for the estimation to performed in a much more computationally efficient manner. For general - and possibly non-separable shapes -, the bispectrum may be decomposed into a sum of separable basis functions (see \cite{Modal10} and references therein, as well as \cite{Regan10} for a similar treatment of the trispectrum);
\ben
(k_1 k_2 k_3)^2 {\rm B}_{\Phi}(k_1,k_2,k_3) = \sum_{n} \alpha_n^Q Q_n(k_1,k_2,k_3)\equiv \sum_{n}\alpha_n^Q [\, q_{n_1}(k_1)q_{n_2}(k_1)q_{n_3}(k_1)+5\;{\rm{cyc.perm.}}\,]\,;
\een  
{where $n$ is an dummy index representing a partial ordering over the triplets $\{n_1,n_2,n_3\}$, which label the one dimensional basis functions $q_{n_1}(k)$ (often chosen to be a polynomial of degree $n_1$), and where the notation $Q_n$ is used as a short hand for the combination in square brackets in the expression on the right hand side.} The CMB bispectrum is then given by
\ben
b_{\ell_1 \ell_2 \ell_3}=f_{\rm{NL}}\sum_n\; \alpha_n^Q \;\int\; dr\; r^2 [\; \tilde{q}^{\ell_1}_{n_1}(r)\tilde{q}^{\ell_2}_{n_2}(r)\tilde{q}^{\ell_3}_{n_3}(r) + \; 5\,{\rm{cyc.perms.}} \; ]\,,
\een
where $\tilde{q}^{\ell_1}_{n_1}(r)=(2/\pi)\int dk k^2 q_{n_1}(k) \Delta_\ell(k) j_\ell(k r)$. Alternatively the CMB bispectrum may be written in the form
\ben
b_{\ell_1 \ell_2 \ell_3}=\sum_n \overline{\alpha}_n^Q Q_n (\ell_1,\ell_2,\ell_3)\,,
\een
where the late-time coefficients $\overline{\alpha}_n^Q$ may be related to the primordial coefficients $\alpha_n^Q$ via a `transfer matrix', $\Gamma_{n m}$, which accounts for the integration over the line of sight (for more details see, for example, \cite{Regan13}), i.e. $\overline{\alpha}_n^Q=\sum_m \Gamma_{n m} \alpha_m^Q$.  This simple prescription allows for primordial models to be efficiently mapped to their CMB counterparts, and for the analysis of non-separable shapes to be performed. This decomposition is utilised in this work for the analysis of the aforementioned models, as well as the non-separable DBI model given by the primordial bispectrum, 
\ben
&& {\rm B}_{\Phi}^{\rm{DBI}}=\frac{1}{(k_1 k_2 k_3)^3 (\sum_i k_i)^2}\left(\sum_i k_i^5 +\sum_{i\neq j}(2k_i^4 k_j -3 k_i^3 k_j^2)+\sum_{i\neq j\neq l}(k_i^3 k_j k_l - 4 k_i^2 k_j^2 k_l )\right)\,,\\
\een
For reference, the flattened bispectrum shape is given by,
\ben
&& {\rm B}_{\Phi}^{\rm{flat}}={1 \over 2}({\rm B}_{\Phi}^{\rm equil}-{\rm B}_{\Phi}^{\rm orthog})\,.
\een 
The different primordial models may give an insight into different microphysical mechanisms at work during the inflationary epoch. As such the standard approach of measuring the bispectrum using a single number, $f_{\rm{NL}}$, appears insufficient. Any possible detection of non-Gaussianity must be accompanied with an analysis of the possible mechanism which may induce it. In this respect the PCA approach described in this paper may prove particularly useful, by identifying the orthogonal directions in the data. Each model may be correlated with each direction in order to identify which model corresponds most with which feature.

We have focussed on the temperature anisotropy, mainly for simplicity. The constraints from the Planck satellite
on $f_{\rm NL}$ are dominated by temperature information and are not expected to improve drastically 
with the inclusion of polarization data.

\subsection{Secondary Non-Gaussianity}
The secondary non-Gaussianities are generated at late time. An important type of
secondary is generated at the level of bispectrum results from cross-talks of
 secondaries such as the Integrated Sachs-Wolfe's (ISW) effect and lensing of CMB by large-scale-structure. 
Secondary non-Gaussianity of a similar form is expected also from coupling of point source (PS)
and lensing as well between the thermal Sunyaev-Zeldovich (tSZ) effect and lensing.

{In the case of the ISW signal, a cross-correlation is expected between the projected lensing potential $\phi$ and the temperature map, resulting in the reduced bispectrum,
\ben 
&& b^{\rm ISW-lens}_{\ell_1\ell_2\ell_3} = -{1 \over 2}\left \{ \myC^{\phi {\rm T}}_{\ell_3} {\cal C}^{\rm TT}_{\ell_1}[{\Pi_{\ell_2} - \Pi_{\ell_1} - \Pi_{\ell_3} }]+ {\rm cyc.perm.}\right \}; \quad\quad \Pi_{\ell} = \ell(\ell+1) \label{eq:define_R}
\een
(see \cite{GoldbergSpergel99a},\cite{GoldbergSpergel99b} for a derivation). In particular, the long wavelength modes of ISW
contribution couples with the short-wavelength modes of fluctuations 
generated due to lensing producing the cross-correlation spectrum, $\myC^{\phi {\rm T}}_{\ell}$, 
between the projected lensing potential $\phi$ and the
secondary contribution.} $\myC^{\rm TT}_{\ell}$ denotes the temperature power-spectrum.
The cross-spectra $\myC^{\phi \rm T}_\ell$ take different forms for ISW-lensing,
PS-lensing or SZ-lensing correlation (\cite{MuHeCoVa11}). The skew-spectrum statistic
has already been applied to WMAP 5 year data release by \cite{Calab10} to probe the correlation of CMB lensing potential and the secondary anisotropies.

The bispectrum for unresolved point sources takes the following form:
\ben
&& b_{\ell_1\ell_2\ell_3}^{\rm PS} = {b}^{\rm PS}.
\label{eq:bispec_intro1}
\een
It is derived assuming the point sources are distributed randomly, {according to a Poisson distribution}. 
The exact value of the parameter ${b}^{\rm PS}\;$ however depends on
the flux limit as well as the mask used in a particular survey. The accuracy of such 
an approximation can indeed be extended by adding contributions from
correlation terms.

The overlap between a secondary source of non-Gaussianity and a primordial source may result in biased estimates for parameters as shall be discussed further in \textsection\ref{sec:joints}.
\section{Optimum Skew-Spectra and related Fisher Matrices} 
\label{sec:skew}
\subsection{Estimation of Individual Skew-Spectra}
Many studies involving primordial non-Gaussianity have used the bispectrum, motivated by the
fact that it contains all the information about $f_{\rm NL}$ (\cite{Babich}). 
It has been extensively
studied (\cite{KSW,Crem03,Crem06,MedeirosContaldi06,Cabella06,Liguori07,SmSeZa09}), with most of these measurements providing 
convolved estimates of the bispectrum. Optimised 3-point
estimators were introduced by \cite{Heav98}, and have been successively developed
(\cite{KSW,Crem06,Crem07b,SmZaDo00,SmZa06}) to the point where an estimator for $f_{\rm NL}$ which saturates
the Cramer-Rao bound exists for partial sky coverage and inhomogeneous noise (\cite{SmSeZa09}). The skew spectrum was devised in \cite{MuHe10} as a method to constrain the bispectrum without reduction to a single parameter.

{The optimum skew-spectrum, $S^{\rm opt}_{\ell}$, associated to a given angular bispectrum, $B$, was introduced in \cite{MuHe10} as a generalisation of the standard KSW estimator \cite{KSW} for non-Gaussianity. The latter reduces the information down to a single amplitude $f_{\rm NL}$, while the former computes a decomposition of the signature {for each angular scale, ${\rm \ell}$}. In particular, accounting for beam and noise effects as described in equation \eqref{eq:beam_noise}), $S^{\rm opt}_{\ell}$ is given by
\ben
S^{\rm opt}_{\ell} = {1 \over 6 }\sum_{\ell_a\ell_b} {{{\rm \tilde{B}}^2_{\ell\ell_a\ell_b}} \over {\tilde{\cal C}_{\ell_a}\tilde{\cal C}_{\ell_b}\tilde{\cal C}_{\ell}}} ={1 \over 6}\sum_{\ell_a\ell_b} {{{\rm {B}}^2_{\ell\ell_a\ell_b}} \over {\myC}^{\rm tot}_{\ell_a}{\myC}^{\rm tot}_{\ell_b}{\myC}^{\rm tot}_{\ell}} \,,
\label{eq:opt}
\een
where our final expression we simplify notation by writing ${\myC}^{\rm tot}_{\ell}=
\tilde{\cal C}_{\ell}/{\cal B}_\ell^2\equiv ({\cal B}_\ell^2 {\cal C}_{\ell}+n_\ell)/{\cal B}_\ell^2$.
One defines the skew-spectrum of the data, $\hat{B}$, as
\ben\label{eq:dataskew}
\hat{S}^{\rm opt}_\ell =  {1 \over 6 \fsky}\sum_{\ell_a\ell_b} {{{\rm {B}}_{\ell\ell_a\ell_b}}{{\rm \hat{B}}_{\ell\ell_a\ell_b}} \over {\myC}^{\rm tot}_{\ell_a}{\myC}^{\rm tot}_{\ell_b}{\myC}^{\rm tot}_{\ell}}\,,
\een
where the parameter $\fsky$ accounts for the effect of partial sky coverage.
Denoting $S^{\rm opt}=\sum_\ell S^{\rm opt}_\ell$ and $\hat{S}^{\rm opt}=\sum_\ell \hat{S}^{\rm opt}_\ell$, we note that the KSW estimator for the amplitude, $\fNL$,  of the bispectrum, $B$, present in the data may be estimated by the quantity $\fNLhat=\hat{S}^{\rm opt}/S^{\rm opt}$. In determining this quantity one should note that the Fisher matrix defined by $\langle \hat{S}^{\rm opt} \hat{S}^{\rm opt}\rangle$ is given by $S^{\rm opt}$, i.e. the estimator is the optimal, inverse Fisher matrix weighted measure of the skewness. In the case of the skew-spectrum one wishes to exploit the extra information available and construct a maximum likelihood estimator at each scale, $\ell$, in the form,
\ben\label{eq:jointx13}
[\hat{f}^{}_{\rm NL}]_{\ell} =  \sum_{\ell'}[{\bf F}^{-1}]^{}_{\ell \ell'} \hat{S}^{}_{\ell'}\,, 
\een
where the Fisher matrix, defined as $[{\bf F}^{}]_{\ell\ell'}\equiv \la\hat S_{\ell}^{^{\rm opt}}\hat S_{\ell'}^{\rm opt}\ra $, is given by the expression
\ben
&&[{\bf F}^{}]_{\ell\ell'}= 
{1 \over 18} \delta_{\ell\ell'} \sum_{\ell_a\ell_b}{{\rm B}^2_{\ell\ell_a\ell_b} \over \myC^{\rm tot}_{\ell}
\myC^{\rm tot}_{\ell_a}\myC^{\rm tot}_{\ell_b}}
+ {1 \over 9 }\sum_{l_a}{{\rm B}^2_{\ell\ell'\ell_a} \over \myC^{\rm tot}_{\ell}\myC^{\rm tot}_{\ell'}\myC^{\rm tot}_{\ell_a}}
  = {1 \over 3} \delta_{\ell\ell'}S^{\rm opt}_{\ell}+ {1 \over 9 } \sum_{l_a}{{\rm B}^2_{\ell\ell'\ell_a} \over \myC^{\rm tot}_{\ell}\myC^{\rm tot}_{\ell'}\myC^{\rm tot}_{\ell_a}}.\label{eq:covopt}
\een
We note reassuringly that, as expected, $\sum_{\ell \ell'}[{\bf F}^{}]_{\ell\ell'}=S^{\rm opt}$.
We have neglected to account for the effect of partial sky coverage in our discussion to here. 
In what follows we shall set the parameter $\fsky$ to unity for simplicity. Nevertheless, in our numerical estimations, we shall use a realistic value of $\fsky$ to account for the effect of incomplete sky coverage.
  }
  
{In the case of secondary models, one may object to use of the parameter $\fNL$. Rather for those models one should consider the parameter $\fNL$ (or more particularly its inverse) as identifying the signal-to-noise of the associated secondary model. Nevertheless, we will persist with these definitions, understanding the parameters to be identified by the expressions given in this section.\footnote{We will make an exception in \textsection\ref{sec:results} in the case of point sources, whose  amplitude we wish to determine from the data and identify $\fNL$ with $b^{\rm{PS}}/10^{-29}$.}}

\subsection{Joint Estimation of multiple skew-spectra}\label{sec:joints}
{
 We next consider the problem of simultaneous estimation of the multiple amplitudes $f_{\rm NL}$ from a given dataset. 
We will assume that the total non-Gaussianity is a sum of contributions from individual components.
\ben
&& {\rm B}_{\ell_1\ell_2\ell_3} = \sum_{\rm X} f^{\rm (X)}_{\rm NL} [{\rm B}^{\rm (X)}]_{\ell_1\ell_2\ell_3}\,.
\een
Here we defined a ${\rm B}^{\rm (X)}_{\ell_1\ell_2\ell_3} = f^{\rm (X)}_{\rm NL}[{\rm B}^{\rm (X)}]_{\ell_1\ell_2\ell_3}$ where $[{\rm B}^{\rm (X)}]_{\ell_1\ell_2\ell_3}$
is the bispectrum of type $(\rm X)$ evaluated at $f_{\rm NL}=1$. The skew spectrum for each model, $\rm X$, given by the data is defined, as usual, by equation \eqref{eq:dataskew}, with $B$ replaced by $[{\rm B}^{\rm (X)}]$ and is denoted $\hat{S}^{(X)}_\ell$. In order to perform a joint estimation of the parameters $\fNL$ associated to two models, ${\rm X}$ and $\rm Y$, one must calculate the Fisher matrix $[{\bf F}]^{(\rm X Y)}=\la \hat S^{({\rm X})}\hat S^{(\rm Y)}\ra$. Correspondingly, in order to carry out the joint estimation of two-different types of skew-spectrum of type ${\rm X}$ and $\rm Y$ it is necessary to calculate the (block) Fisher matrix, $[{\bf F}]^{(\rm X Y)}_{\ell \ell'}$, with:
\ben
&&[{\bf F}]^{(\rm X Y)}_{\ell \ell'}= \la\hat S_{\ell}^{({\rm X})}\hat S_{\ell'}^{({\rm Y})}\ra = 
{1 \over 18} \delta_{\ell\ell'} \sum_{\ell_a\ell_b}{{\rm B}_{\ell\ell_a\ell_b}^{(\rm X)}{\rm B}_{\ell\ell_a\ell_b}^{(\rm Y)} \over \myC^{\rm tot}_{\ell}
\myC^{\rm tot}_{\ell_a}\myC^{\rm tot}_{\ell_b}}
+ {1 \over 9}\sum_{l_a}{{\rm B}_{\ell\ell'\ell_a}^{(\rm X)}{\rm B}_{\ell\ell'\ell_a}^{(\rm Y)} \over \myC^{\rm tot}_{\ell}\myC^{\rm tot}_{\ell'}\myC^{\rm tot}_{\ell_a}}   = {1 \over 3} \delta_{\ell\ell'}S^{\rm opt}_{\ell}({\rm X,Y})+ {1 \over 9} \sum_{l_a}{{\rm B}^{(\rm X)}_{\ell\ell'\ell_a}
{\rm B}^{(\rm Y)}_{\ell\ell'\ell_a}  \over \myC^{\rm tot}_{\ell}\myC^{\rm tot}_{\ell'}\myC^{\rm tot}_{\ell_a}}\,,\label{eq:covopt1}
\een
where we have introduced the following notation above:
\ben
&&S^{\rm opt}_{\ell}({\rm X,Y}) \equiv \frac{1}{6} \sum_{\ell_a\ell_b}{{\rm B}_{\ell\ell_a\ell_b}^{(\rm X)}{\rm B}_{\ell\ell_a\ell_b}^{(\rm Y)} \over \myC^{\rm tot}_{\ell}
\myC^{\rm tot}_{\ell_a}\myC^{\rm tot}_{\ell_b}}\,,
\label{eq:skewspec}
\een
and similarly for $S^{\rm opt}_{\ell}({\rm X,X})$ and $S^{\rm opt}_{\ell}({\rm Y,Y})$. {The Fisher matrix is understood to be in (symmetric) block form with each block corresponding to a specific combination of models $({\rm X,Y})$. Within each block, $({\rm X,Y})$, the $(\ell,\ell')$ components are given by $[{\bf F}]^{(\rm X Y)}_{\ell \ell'}$.} We note again that 
\ben
&&[{\bf F}]^{(\rm X Y)}=\sum_{\ell \ell'} [{\bf F}]^{(\rm X Y)}_{\ell \ell'}=\sum_\ell S^{\rm opt}_{\ell}({\rm X,Y}) \equiv {1 \over 6}\sum_{\ell_1\ell_2\ell_3} {[{\rm B}]^{\rm(X)}_{\ell_1\ell_2\ell_3}[{\rm B}]^{\rm(Y)}_{\ell_1\ell_2\ell_3}\over \myC_{\ell_1}^{\rm tot}\myC_{\ell_2}^{\rm tot}\myC_{\ell_3}^{\rm tot}  } \,,\label{eq:FisherA}
\een
}
The inverse Fisher matrix $[{\bf F}^{-1}]^{\rm(XY)}$ defines error-covariance
matrix for the $f_{\rm NL}$ parameters,  
\ben\label{eq:fisherXY}
&& [{\bf F}^{-1}]^{\rm(XY)} = \la \delta f^{(\rm X)}_{\rm NL}\delta f^{(\rm Y)}_{\rm NL} \ra \,,
\een
with the expected error bar for each parameter $f_{\rm NL}$ given by $\delta f_{\rm NL}^{(\rm X)}=\sqrt{[{\bf F}^{-1}]^{\rm(XX)}}$.
One may also define a correlation measure between models $(\rm X)$ and $(\rm Y)$ given by {${\rm{r}}({\rm X,Y})=[{\bf F}]^{\rm(XY)}/\sqrt{[{\bf F}]^{\rm(XX)}[{\bf F}]^{\rm(YY)}} $}. 
The maximum likelihood estimator may be written in terms of the inverse Fisher matrix as:
\ben\label{eq:joint1}
&&\hat{f}^{\rm(X)}_{\rm NL} = \sum_{\rm Y} [{\bf F}^{-1}]^{\rm(XY)} \hat{S}^{\rm(Y)}\,; \quad\quad\hat S^{(\rm Y)}= {1 \over 6}\sum_{\ell_1\ell_2\ell_3} {[{\rm B}]^{\rm(Y)}_{\ell_1\ell_2\ell_3}{\hat {\rm B}}_{\ell_1\ell_2\ell_3}\over \myC_{\ell_1}^{\rm tot}\myC_{\ell_2}^{\rm tot}\myC_{\ell_3}^{\rm tot}  }\,.
\een
This formalism is easily extended to the case of the skew-spectrum with the Fisher matrix for the cross-spectra, $[{\rm F}^{(\rm XY)}]_{\ell \ell'}$ in this case given by equation~\eqref{eq:covopt1}. 
The maximum likelihood estimator at each scale, $\ell$, may be expressed in the form,
\ben\label{eq:joint3}
&&[\hat{f}^{\rm(X)}_{\rm NL}]_{\ell} = \sum_Y \sum_{\ell'}[{\bf F}^{-1}]^{\rm(XY)}_{\ell \ell'} \hat{S}^{\rm(Y)}_{\ell'};
\quad\quad
\hat S^{(\rm Y)}_{\ell} = {1 \over 6}\sum_{\ell_2\ell_3} {[{\rm B}]^{\rm(Y)}_{\ell \ell_2\ell_3}{\hat{\rm B}}_{\ell \ell_2\ell_3}\over \myC_{\ell }^{\rm tot}\myC_{\ell_2}^{\rm tot}\myC_{\ell_3}^{\rm tot}  }\,. 
\een
The inversion of the block matrix $[{\bf F}]^{\rm X Y}_{\ell\ell'}$ becomes very numerically challenging for large numbers of models. Instead in this paper we will restrict to the case of using observations of the skew-spectrum, $\hat{S}_l^{(\rm Y)}$, for a {\it{single model}} ${\rm Y}$, in order to infer the quantity $[\hat{f}_{\rm NL}^{(\rm X)}]_\ell$ for (a possibly distinct) model $\rm X$. However, as we shall see, for the case of the PCA components, the Fisher matrix becomes block diagonal, making inversion relatively trivial.

{The presence of a secondary bispectrum may induce a non-zero value of the estimator of a primordial model due to a non-zero overlap between the two bispectra. This value is termed the $bias$ and must be corrected for in measurements of the estimator, $f_{\rm{NL}}$.}
{More concretely,} the expected bias for a primordial model $(\cal P)$ due to a secondary model $(\cal S)$ is given by
\ben
&&\delta_b f_{\rm{NL}}^{\cal P} =  [{\bf F}^{-1}]^{\cal(P P)}{ [{\bf F}]^{\cal(P S)} }{   }\,.\label{eq:bias}
\een
A joint estimation of primary and secondary non-Gaussianity allows us to marginalise over the presence of secondaries in order to provide accurate estimates for the bias as well as their impact on the error bars.
\begin{figure}
\centering
\large{\textrm{\;\;\;\;Fisher Matrices for Primary and Secondary Skew-Spectra}}\par
\vspace{0.25cm}
\hspace{0.1cm}
{\includegraphics[width=0.32\linewidth]{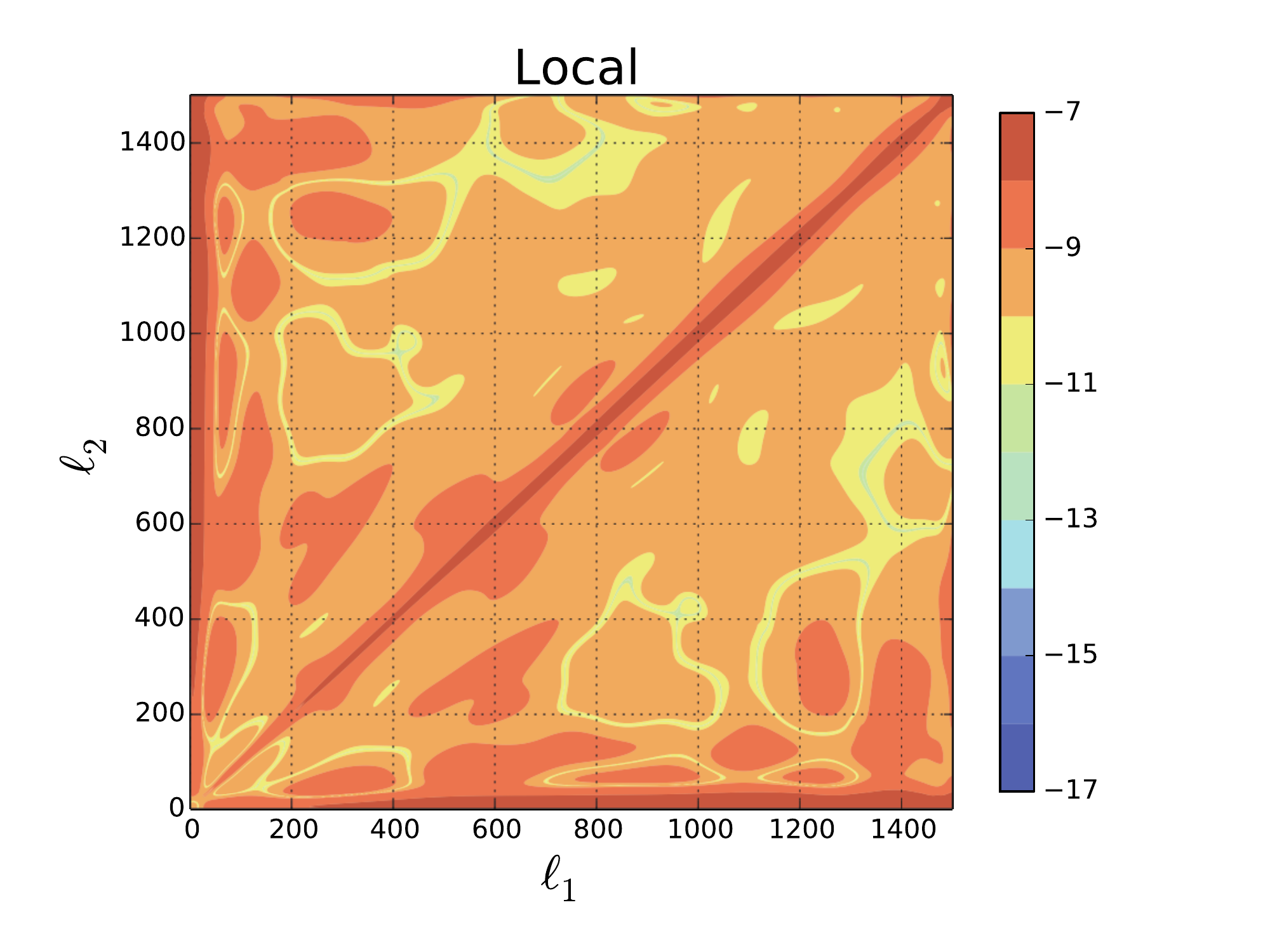}}
\hspace{0cm}
{\includegraphics[width=0.32\linewidth]{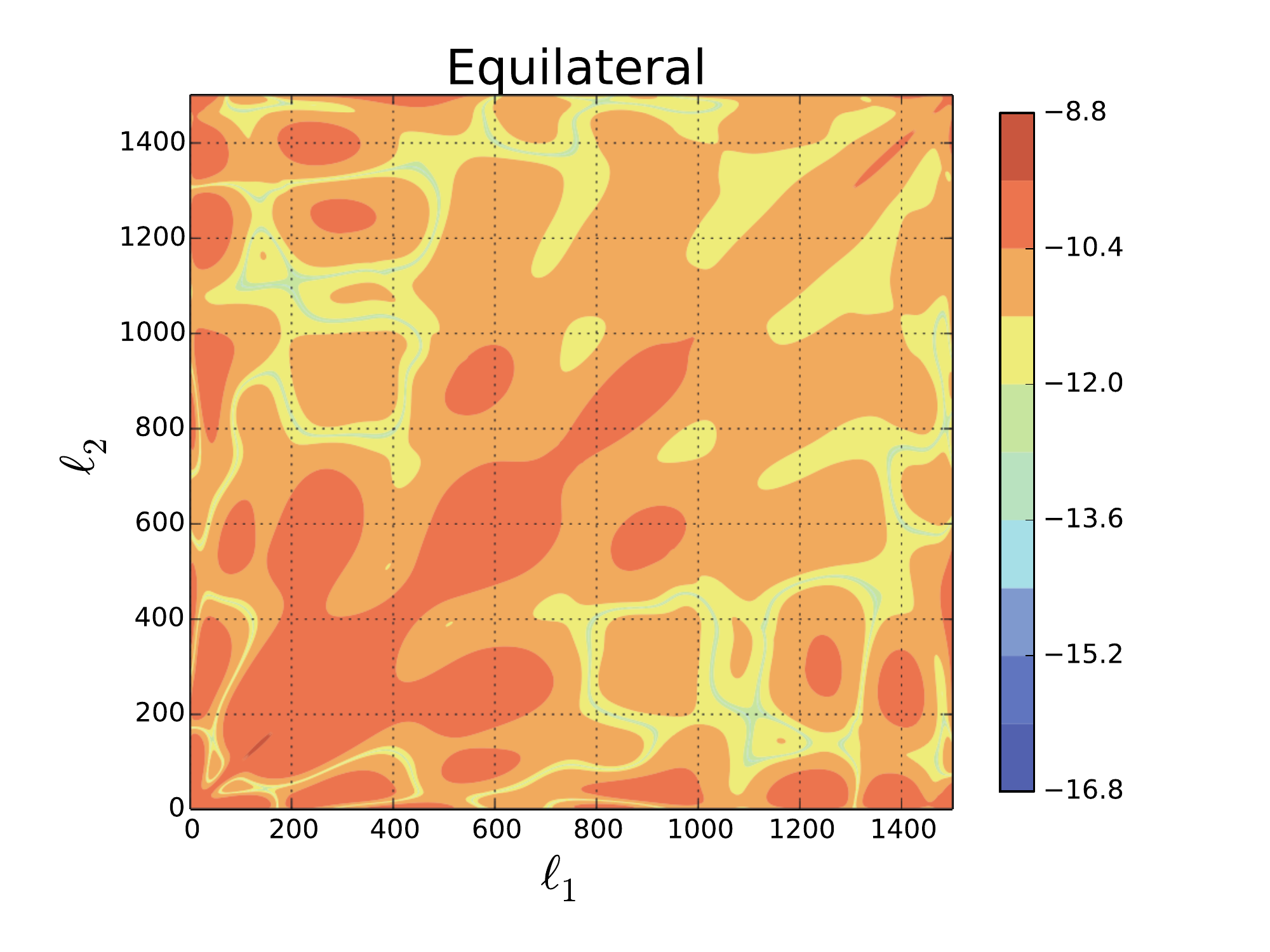}}
{\includegraphics[width=0.32\linewidth]{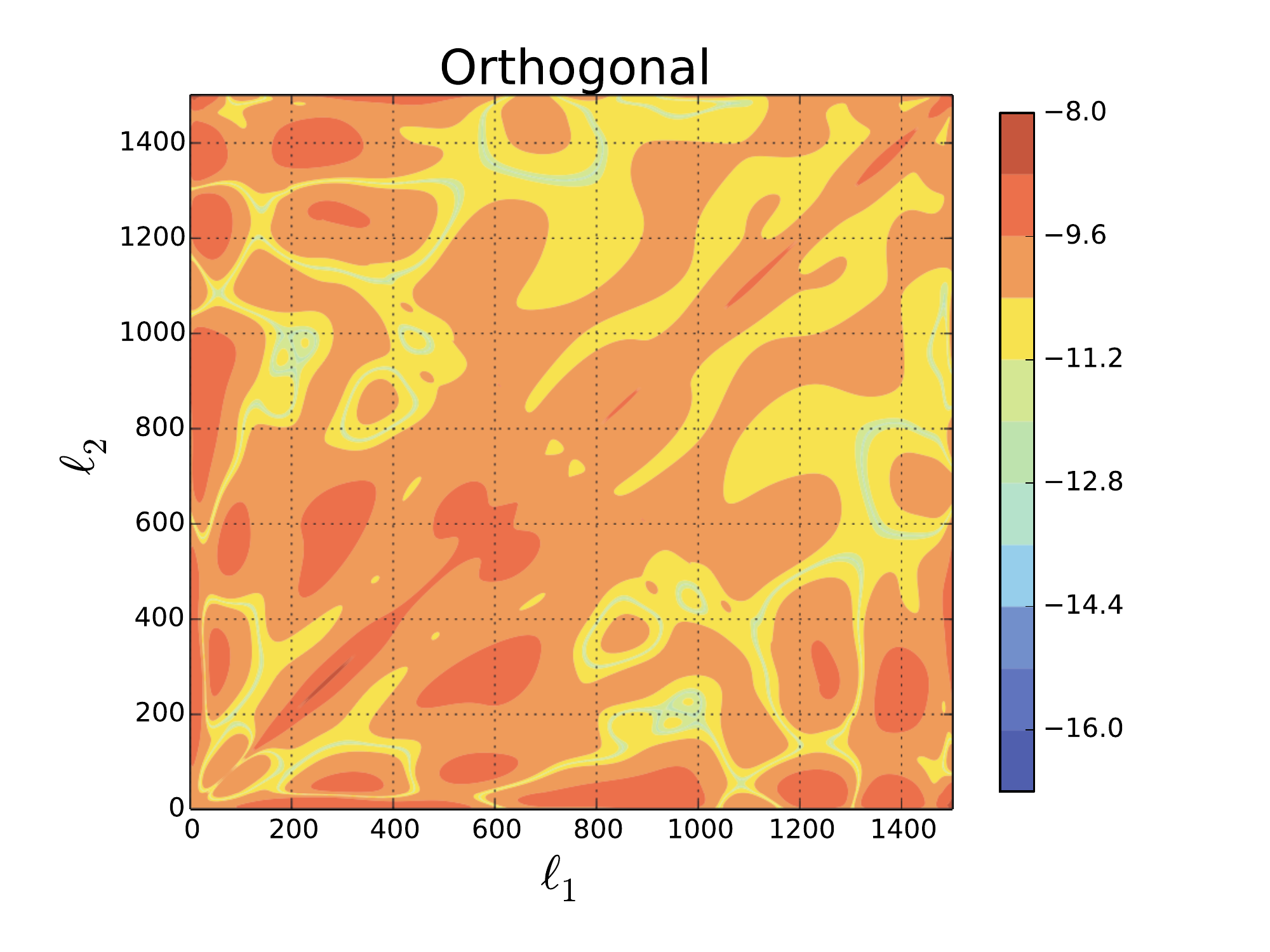}} \\
{\includegraphics[width=0.32\linewidth]{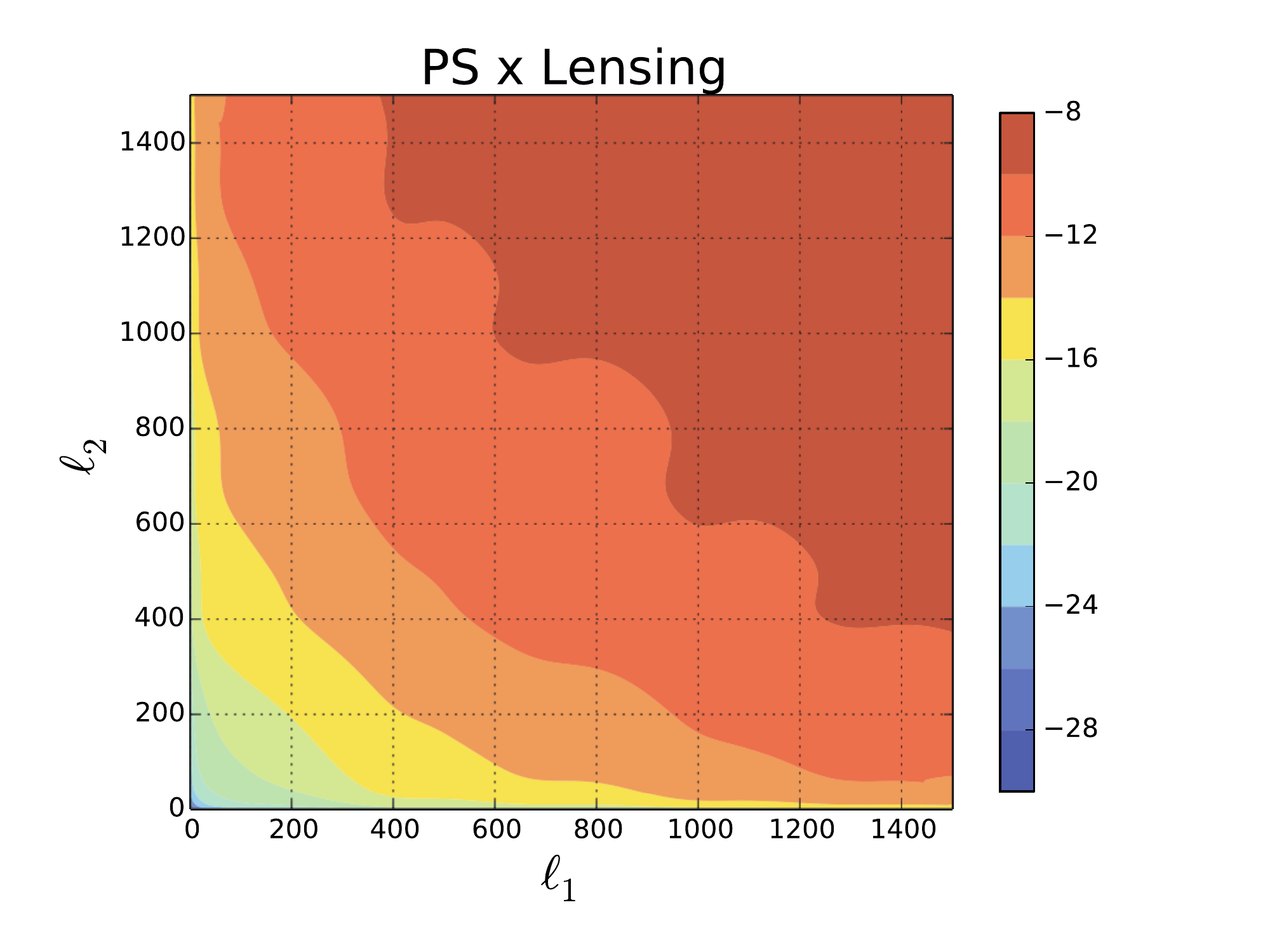}}
\hspace{0cm}
{\includegraphics[width=0.32\linewidth]{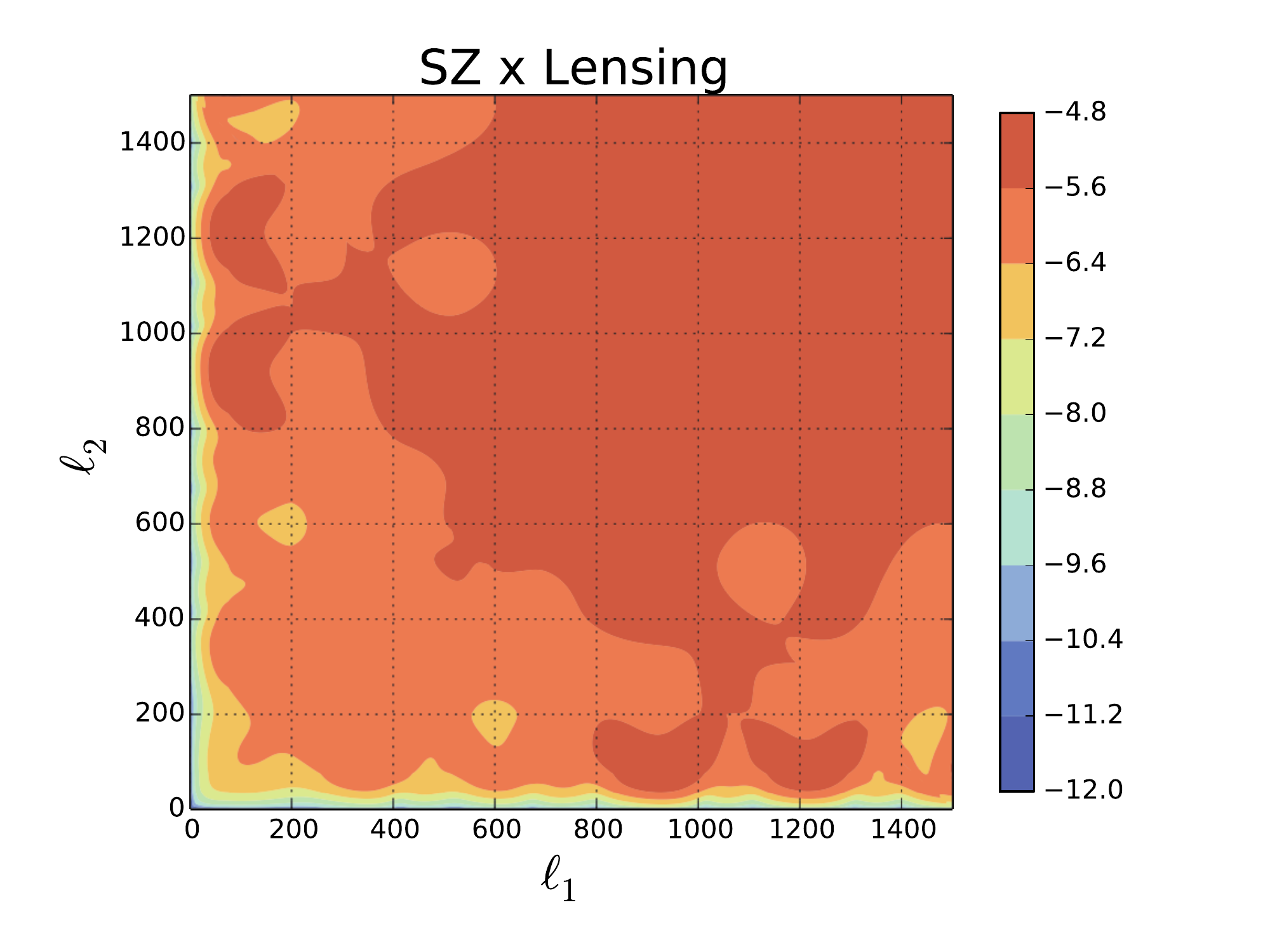}}
{\includegraphics[width=0.32\linewidth]{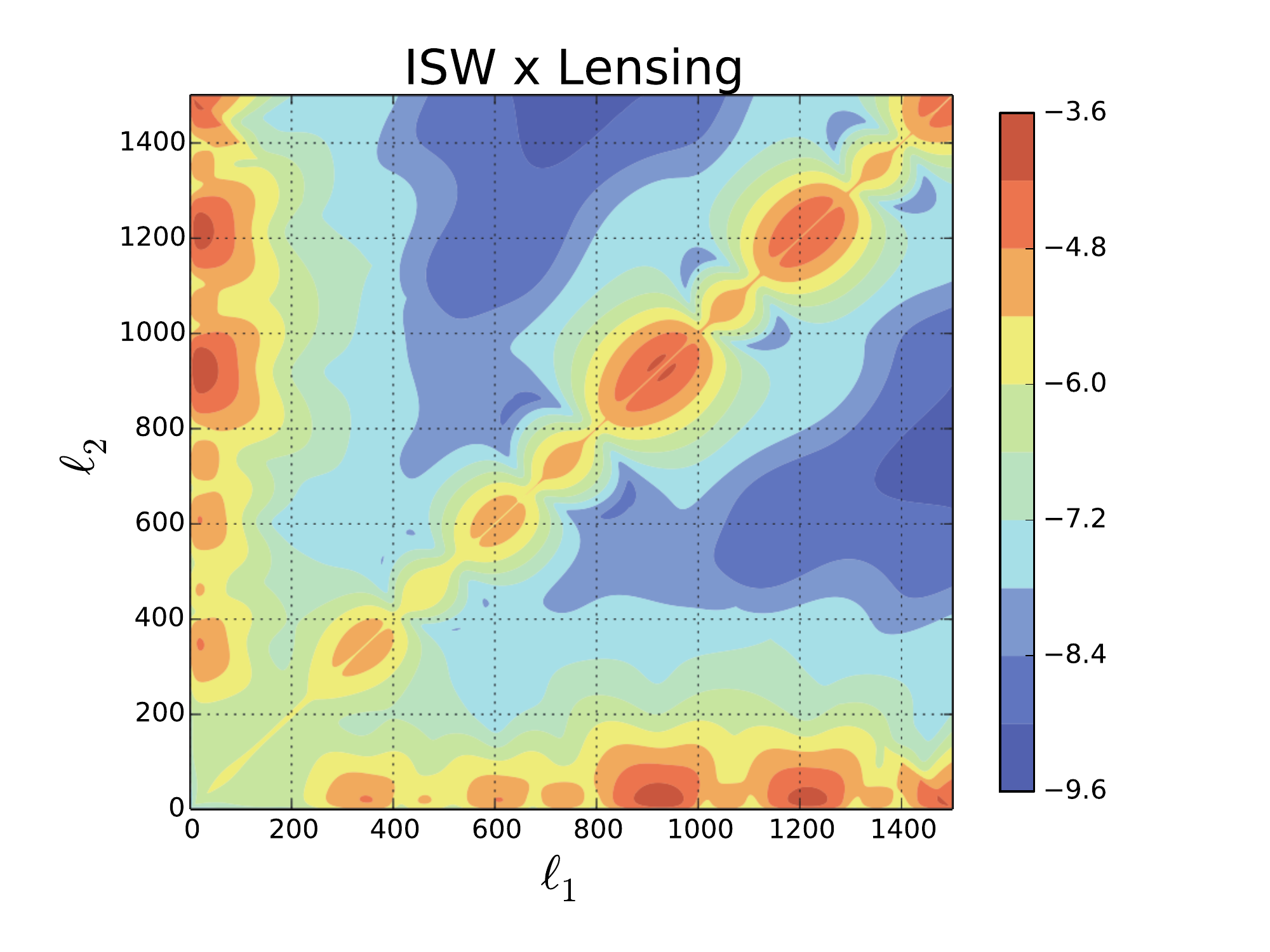}}\\
\caption{The ``{\em off-diagonal}" elements of the Fisher matrix associated with the optimum skew spectra, $S_{\ell}^{\rm{opt}}$ for selection of primordial (local, equilateral and orthogonal) and secondary bispectra (ISW x Lens, SZ x Lens and residual point sources) are displayed. The expression for the Fisher matrix is given in 
equation (\ref{eq:covopt}) (the ``{\em off-diagonal}" elements refer to contribution due solely to the second term in that equation). The optimum skewspectrum is defined in equation (\ref{eq:opt}). The top row corresponds to models for primordial non-Gaussianity and
the bottom row correspond to secondary non\_Gaussianity. We use a Planck-realistic experiment at frequency 143GHz;
we use the parameters $\theta_b = 7.1'$ , $\sigma_{\rm pix} = 2.2 \times 10^{-6}$ and
$\Omega_{\rm pix} = 0.0349$. We perform our computations out to $\ell_{\rm max} = 1500$, beyond which the signal is increasingly noise dominated
(see text for more details).}
\label{fig:skewspec}
\end{figure}

\section{Truths and Myths Concerning Principal Component Analysis}
\label{sec:principal}
{
\subsection{General Considerations}
Given the number and variation of primary and secondary models of non-Gaussianity studied in the literature, it is clearly desirable to perform an estimation of each model in a timely fashion. Rather than re-implementing the full estimator for each model individually, one may instead wish to compress the available information in the dataset. In particular, if one has in mind a particular model which overlaps strongly with an already constrained template, one may wish to use existing constraints on the template to infer constraints on the model of interest. A principal component analysis (PCA) formalises this procedure. In this paper we will describe $two$ PCA procedures, applied to the joint analysis of various models, $\{{\rm X}\}$ analysed using their skew statistic, $S^{({\rm X})}$ to identify orthogonal models amongst the set $\{{\rm X}\}$, and an analysis of the skew-spectrum of individual models $S^{(X)}_\ell$, in order to identify the independent information contained in the spectral decomposition of the skewness encapsulated in this statistic. In this section we will firstly describe the PCA procedure on general grounds, and then specify to the skew-statistic.}
\par
{ Given a set of parameters, $\{\Theta_i\}$, ({in this paper the parameters correspond to the skew-spectra}) the Fisher matrix is defined as ${\bf{F}}_{ij}=\langle \delta\Theta_i \delta\Theta_j\rangle$, where $\delta \Theta_i=\Theta_i-\la\Theta_i\ra$, i.e. $\delta \Theta_i$ represents the difference between the parameter and its expectation value. Generally the parameters will be correlated, i.e. ${\bf{F}}_{ij}\neq 0$ for $i\neq j$. In order to identify the orthogonal directions in parameter space, one may diagonalise the (symmetric) Fisher matrix in the form\footnote{We will use Einstein summation convention throughout this section, unless otherwise specified.}
\ben
&&{\rm F}_{ij} = {\rm W}_{ik}^{\rm T} {\Lambda}_{kl} {\rm W}_{lj}.
\label{eq:window}
\een
The choice of {\bf W} is not unique. Any orthogonal rotation ${\bf OW}$ with
${\bf O}\in {\bf SO}(n)$ is also valid. If W is an orthogonal matrix, its rows are the eigenvectors of $p_i$ of $\bf F$ with ${\bf \Lambda}={\rm{diag}}(\lambda_i)$ labelling the corresponding eigenvalues. In this case ${\bf F} = {\bf W}^{\rm T} \Lambda {\bf W}$ is called the principal component decomposition. We will, therefore, assume that ${\bf W}$ is orthogonal, i.e.$ {\bf W}_{i j}{\bf W}^T_{j k}=\delta_{i k}$.
\\
Defining the quantities $\delta \Phi_i = W_{i j}\delta \Theta_j$, we find that the corresponding Fisher matrix is diagonal \cite{HT00}:
\ben
&&\la\delta\Phi_i\delta\Phi_j\ra = {\rm W}_{i k} \langle \delta\Theta_k \delta\Theta_l \rangle {\rm W}_{l j}^{\rm T} = \Lambda_{ij}=\lambda_i \delta_{ij}\,. 
\een
Therefore the parameters  $\delta \Phi_i$ are orthogonal directions in parameter space.
The eigenvectors or the principal components of ${\bf F}$ determine the principal axes of the n-dimensional error
ellipsoid in parameter space. We will introduce the notation for the principal directions $P_i=\delta \Phi_i=W_{i j}\delta \Theta_j$.
The accuracy with which these parameters can be determined is quantified by the variance (since the corresponding Fisher matrix is diagonal)
$\sigma_i \equiv \sigma({\rm P}_i)= 1/\sqrt{\la (\delta {\rm \Phi}_i)^2 \ra} = \lambda^{-1/2}_{i}$. The eigenvalues are sorted
in descending order, with the first eigenvector ${\rm P}_i$ having the smallest variance, corresponding 
to the best constrained parameter
combination. The last eigenvector ${\rm P}_n$ is the direction with the largest uncertainty. 
Using equation (\ref{eq:window}) we can reconstruct the errors of individual physical parameters:
\ben
&&\delta \Theta_i = \left [ \sum_{j=1}^n {\rm W}^2_{ji}/\lambda_j \right ]^{1/2}.
\een 
While naively the dimension of the rotated parameter space, $\{P_i\}$ is the same as that of the original parameters $\{\Theta_i\}$, one may find that those eigenvectors with the largest uncertainty contribute little to the constraining power of particular models, and therefore may be neglected. Therefore, PCA formalises the procedure of reducing the dimensionality of the parameter space of interest. 
\\
Given the analysis of several models, one may identify the principal components. These orthogonal modes, can then be used to constrain a separate model, by correlating this model to the principal components. This has the potential for greatly reducing the numerical effort involved in constraining models. We will describe this procedure in \textsection\ref{sec:bispecPCA} and {illustrate the technique using the example} of the DBI bispectrum in \textsection\ref{sec:results}. As we will describe, the DBI model may be constrained using standard templates (through the use of a simple look-up table) by the use of a simple correlation measure and an estimate of the Fisher bound, and without the need to reproduce the CMB analysis pipeline. This procedure may be replicated for any (smooth) model.\footnote{The condition of smoothness is such that the model under consideration lies within the parameter space sampled by the standard templates.}
}
\begin{figure}
\centering
\large{\textrm{\;\;\;\;Primordial and Secondary Skew-Spectra}}\par
\vspace{-0.1cm}
\hspace{0.1cm}
{\includegraphics[width=0.8\linewidth]{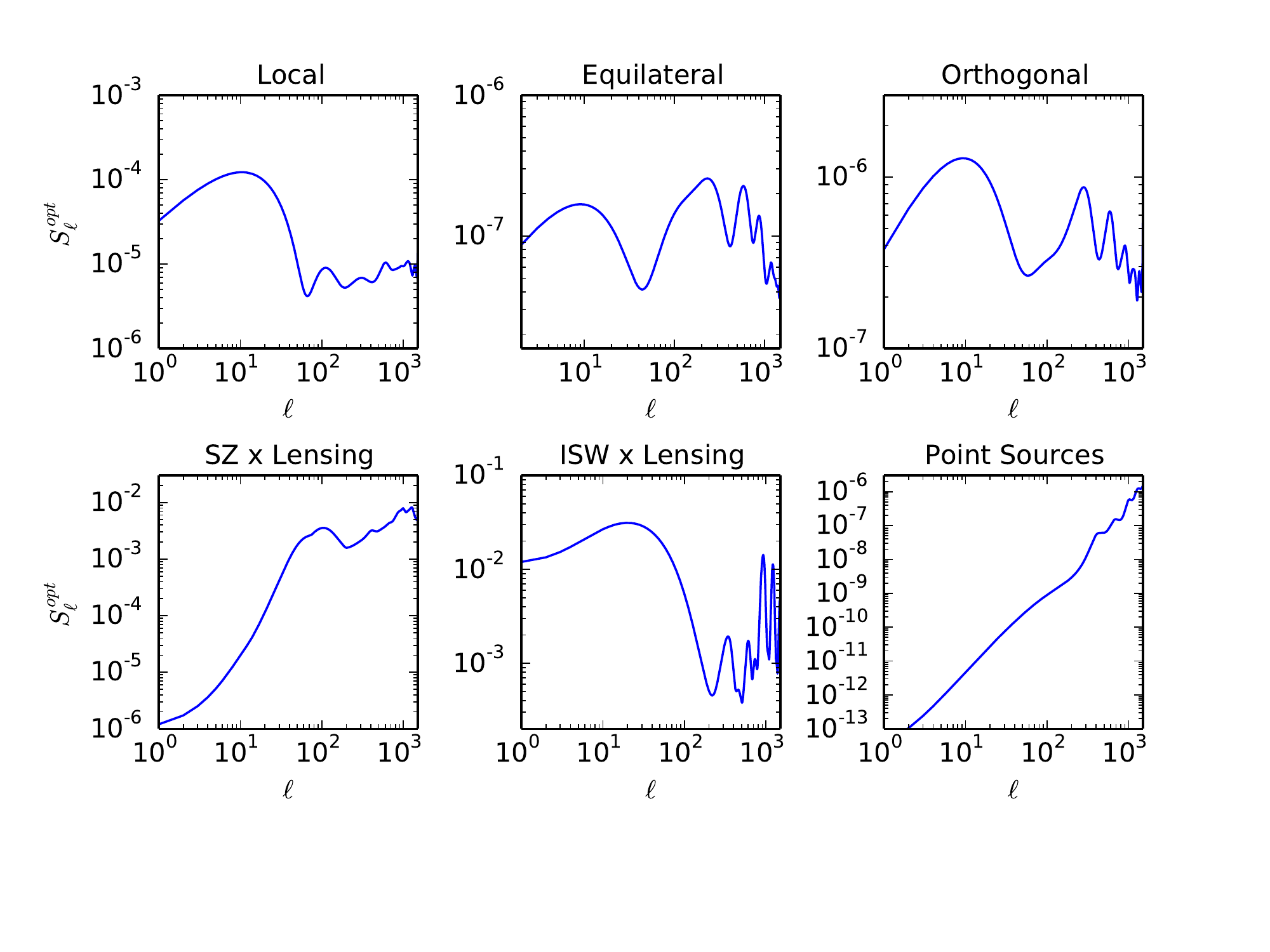}}
\vspace{-1.2cm}
\caption{The skew-spectra associated with primordial (top panels) and secondary (bottom panels) non-Gaussianities are depicted.
The skew-spectra is defined in equation (\ref{eq:opt}). A Planck-realistic experiment at frequency 143GHz;
was used with parameters $\theta_b = 7.1'$ , $\sigma_{\rm pix} = 2.2 \times 10^{-ˆ'6}$ and
$\Omega_{\rm pix} = 0.0349$.}
\label{fig:skewspec2}
\end{figure}
{
\subsection{PCA analysis of the bispectrum}\label{sec:bispecPCA}
}
More concretely, for a joint analysis of multiple bispectra, $\{{\rm B}_i\}$, the $(i,j)$ Fisher entry is the overlap between models $i$ and $j$, i.e. ${\rm F}_{ij}=\langle {\rm B}_i {\rm B}_j\rangle$, {given in the case of the skew-spectrum by equation~\eqref{eq:FisherA}, i.e. ${\rm F}_{ij}\equiv \langle {\hat S}^{(i)}  {\hat S}^{(j)}\rangle$}.  Constructing the combination ${\rm P}_i={\rm W}_{i j}{\rm B}_j$ we form the orthogonal (or decorrelated) combinations with Fisher matrix $[{\bf F}]^{{\rm P}_i {\rm P}_j}=\lambda_i \delta_{ij}$.

While Planck has produced constraints on the range of primordial models listed already, one may wish to estimate the error bars and expected bias for another model, ${\rm M}$. Rather than redo the entire analysis for this particular model, one can perform an approximate (but generally quite accurate) analysis in the following way: 
\begin{itemize}
\item Using a primordial measure between two bispectra, ${\rm B}^{(\rm X)}_{\Phi}$ and ${\rm B}^{(\rm Y)}_{\Phi}$, 
$\llangle {\rm B}^{(\rm X)}_{\Phi} {\rm B}^{(\rm Y)}_{\Phi}\rrangle$ 
\footnote{For example, one might use the measure $\llangle {\rm B}^{(\rm X)}_{\Phi} {\rm B}^{(\rm Y)}_{\Phi}\rrangle = \int_{2\rm{max}\{k_i\} \leq \sum k_i} dk_1 dk_2 dk_3 S^{(\rm X)}_{\Phi}(k_1,k_2,k_3)S^{(\rm Y)}_{\Phi}(k_1,k_2,k_3)$, where $S^{(\rm X)}_{\Phi}=(k_1 k_2 k_3)^2 
B^{(\rm X)}_{\Phi}$\,.}, {the correlation ${\rm{r}}({\rm X,Y})$ between two models (defined under equation~\eqref{eq:fisherXY}) may be approximated as 
\ben
{\rm{r}}(\rm X,Y)\approx {\rm{r}}_{p}(\rm X,Y)\equiv\llangle {\rm B}^{(\rm X)}_{\Phi} {\rm B}^{(\rm Y)}_{\Phi}\rrangle/\sqrt{\llangle {\rm B}^{(\rm X)}_{\Phi} {\rm B}^{(\rm X)}_{\Phi}\rrangle\llangle {\rm B}^{(\rm Y)}_{\Phi} {\rm B}^{(\rm Y)}_{\Phi}\rrangle}\,.
\een
}
 In this manner one may find the correlation between the model under consideration, $\rm M$, and the set of templates already constrained, $\{{\rm B}_i\}$. {The advantage of the primordial correlation measure, ${\rm{r}}_{p}(\rm X,Y)$, is that it is computationally much more efficient than the CMB 
 correlation measure, ${\rm{r}}_{}(\rm X,Y)$, with the latter requiring the computation of the CMB bispectrum corresponding to each primordial shape. 
 In \cite{ARS14} the accuracy of this approximation was established in the context of the principal components of effective single field models of inflation. For a more detailed discussion of the primordial measure we refer the reader to \cite{Modal10}. }
\item From the set of templates $\{{\rm B}_i\}$ we consider the set of orthogonal shapes ${\rm P}_i=\sum_j{\rm W}_{i j}{\rm B}_j$, where $\langle {\rm P}_i {\rm P}_j\rangle=\lambda_i \delta_{ij}$. We may express model $\rm M$ in terms of these orthogonal shapes in the form 
\ben\label{eq:MbyPCA}
{\rm M}\approx\sum_i {\rm A}_i {\rm P}_i, {\rm where\,\, } 
{\rm A}_i=\langle {\rm M}\,{\rm P}_i\rangle/\langle {\rm P}_i {\rm P}_i\rangle\equiv {1 \over \lambda_i}[{\bf F}]^{{\rm M} {\rm P}_i}.
\een
 The accuracy of this approximation is dependent on the set templates $\{{\rm B}_i\}$ forming a complete basis. We may further simplify and use the approximation for the correlation measure to express, ${\rm A}_i= ({[{\bf F}]^{\rm M M}/\lambda_i})^{1/2}{\rm{r}}({\rm M},{\rm P}_i)\approx  
({[{\bf F}]^{\rm M M}/\lambda_i})^{1/2}{\rm{r}_p}({\rm M},{\rm P}_i)$.  
\item The maximum likelihood estimator given by equation~\eqref{eq:joint1} may then be expressed solely in terms of the estimators for the principal components, with 
\ben
\hat{f}_{\rm NL}^{(\rm M)}\approx \left [ \sum_i \rm A_i \hat{\rm S}^{(\rm P_i)} \right ] \left [\delta f_{\rm NL}^{(\rm M)}\right ]^2 
\equiv \left [ \sum_i \sum_j {\rm A}_i {\rm W}_{i j}\hat{\rm S}^{({\rm B}_j)} \right ]\left [\delta f_{\rm NL}^{(\rm M)} \right ]^2 \, .
\een
\item The error bar for the model may also be approximated using the principal components by 
$\delta f_{\rm NL}^{\rm M}=({[{\bf F}]^{\rm MM}})^{1/2}\equiv({\langle \rm M \rm M\rangle})^{-1/2}\approx (\sum_i {\rm A}_i^2 \lambda_i)^{1/2}$, while the expected bias due to a secondary model, $\cal S$, given by 
equation (\ref{eq:bias}), may be approximated as
$\delta_b f_{\rm NL}^{\rm M} = \sum_{i j} {\rm A}_i {\rm W}_{i j}\langle {\rm B}_j S \rangle/({\sum_i {\rm A}_i^2 \lambda_i})$\,.
\end{itemize}
\par
As described in \textsection\ref{sec:joints}, performing a full Fisher analysis of the skew-spectra of several models becomes numerically very challenging for large numbers of modes. However, one may use the simpler analysis for joint estimation, describing the models using their skewness parameters (c.f. equation (\ref{eq:joint1})), in order to identify the orthogonal shapes, ${\rm P}_i={\rm W}_{i j}{\rm B}_j$. Considering these orthogonal combinations, one need only consider the Fisher matrix, $[{\bf F}]^{{\rm P}_i {\rm P}_i}_{\ell \ell'}$, (since $[{\bf F}]^{{\rm P}_i {\rm P}_j}_{\ell \ell'}$ may be set to zero for $i\neq j$). One then computes the quantities (c.f. equation (\ref{eq:joint3}))
\ben
[f_{\rm NL}^{{\rm P}_i}]_\ell = \sum_{\ell'}[{\bf F}^{-1}]^{({\rm P}_i {\rm P}_i)}_{\ell \ell'} \hat{\rm S}^{({\rm P}_i)}_{\ell'}\,.
\een
One may relate the quantities $[f_{\rm NL}^{{\rm P}_i}]_{\ell}$ to individual templates. The advantage of performing a PCA analysis is thus perhaps even more apparent. Having identified the orthogonal directions, a more complete analysis of each orthogonal mode in terms of its skew-spectrum may be performed, taking full advantage of the data available to more clearly identify features that may be present at different scales, $\ell$, therefore allowing for a more exacting analysis of the consistency of the data with particular models.

\section{Results: Implications for Planck-type experiments}\label{sec:results}
We apply the methodology described in the previous sections to the case of Planck-like data. For this purpose we require the noise power spectrum and beam function, as described in \textsection\ref{sec:skew}. As described in \cite{Baumann09} the beam $b_l$ and noise $n_l$ may be characterised by the parameters $\sigma_{\rm beam}$ and $\sigma_{\rm rms}$, respectively, and
\ben
{\cal B}_{\ell}(\theta_b)=\exp(-\ell(\ell+1)\sigma_{\rm beam}^2)\,; \quad \sigma_{\rm beam}=\frac{\theta_b}{\sqrt{8\ln 2}}\,; \quad   n_l=\sigma_{\rm pix}^2 \Omega_{\rm pix}\,; \quad \Omega_{\rm pix}=\frac{4\pi}{{\rm N}_{\rm{pix}}}\,;
\label{eq:beam}
\een
where $\theta_b$ describes the resolution of the beam, ${\rm N}_{\rm pix}$ represents the number of pixels (of area $\Omega_{\rm pix}$) required to cover the sky, and $\sigma^2_{\rm pix}$ describes the variance per pixel. For a Planck-realistic experiment at frequency $143$GHz we use the parameters $\theta_b=7.1'$, $\sigma_{\rm pix}=2.2\times 10^{-6}$ and $\Omega_{\rm pix}=0.0349$. We perform our computations out to  $\ell_{\rm max}=1500$, beyond which the signal is increasingly noise dominated. 

We shall consider in our PCA analysis the three standard templates for the primary sources of non-Gaussianity, i.e. local, equilateral and orthogonal bispectra. We shall, in addition, discuss results applied to the DBI model as an example of the usefulness and efficacy of the PCA analysis described. We also analyse four secondary templates -  three due to the cross correlation of lensing with the ISW, SZ and point source signal, respectively, and the other due to point sources only. With regard to nomenclature we will use the $f_{\rm NL}$ parameter to signify the amplitude of the primary {\it and} secondary signals. For the latter case $\delta f_{\rm NL}$ may be understood as representing the inverse of the signal to noise. In the case of the point source only bispectrum, as given by equation~\eqref{eq:bispec_intro1}, we characterise the amplitude in units of $10^{-29}$, i.e. $ f_{\rm NL}= b^{\rm PS}/10^{-29}$.

\begin{figure}
\centering
\large{\textrm{\;\;\;\;Eigenvalues corresponding to various Skew Spectra}}\par
\vspace{-0.15cm}
{\includegraphics[width=0.9\linewidth]{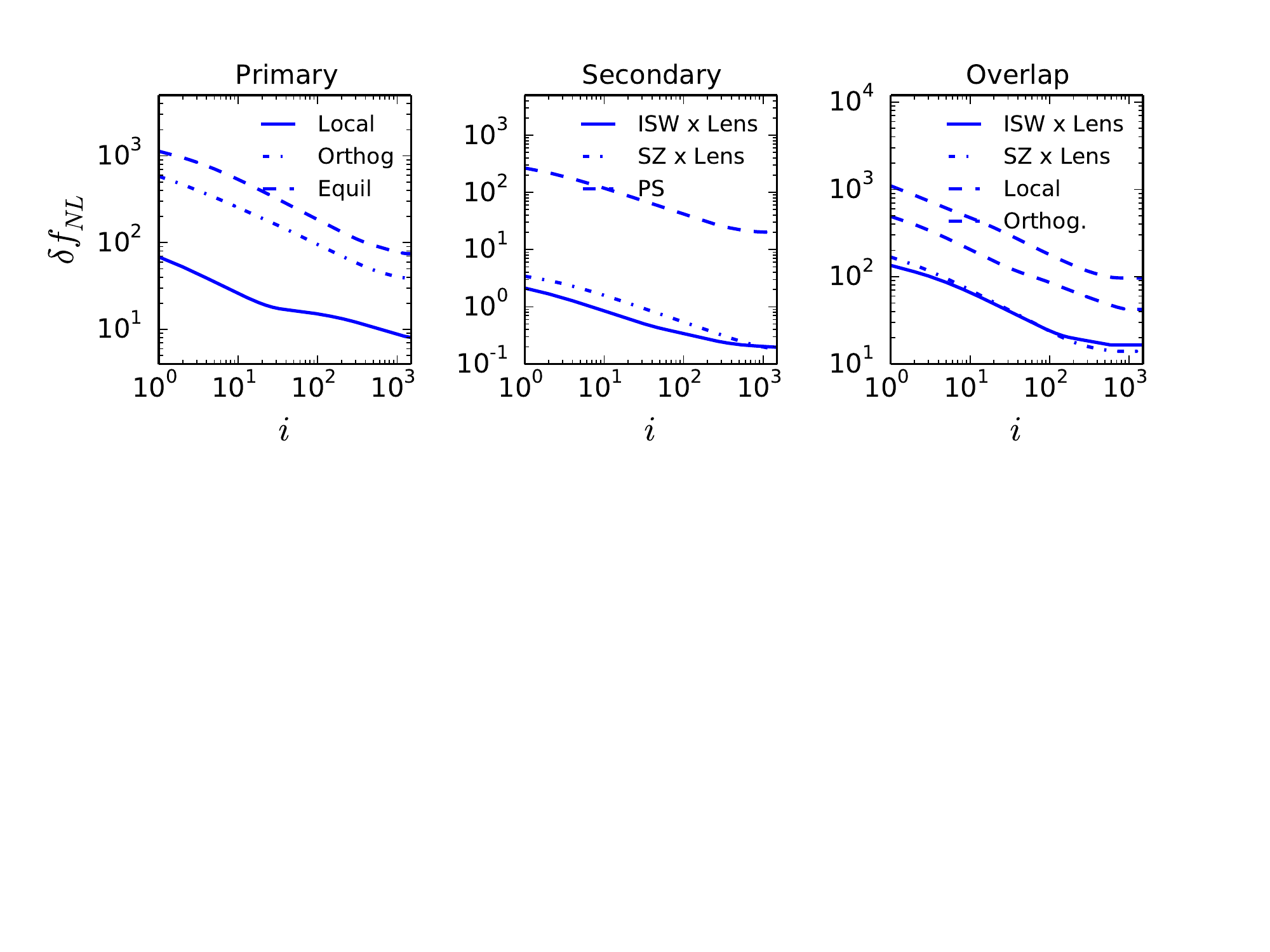}}
\vspace{-5.6cm}
\caption{Plot of $i$ versus $1/\sqrt{3\sum_{n=1}^{i}\lambda_n}$ for the eigenvalues for the primary skew-spectra (left panel), the secondary skew-spectra (middle panel), and the overlap between various models and the equilateral model (right panel) are shown. The sum converges to $\delta f_{NL}$ as we include more eigenvalues.}
\label{fig:eigen}
\end{figure}

\subsection{Results for Individual Skew-Spectra}
\label{sec:indi}
The skew spectrum, $S_l$, is a useful statistic with which to estimate the signal from a primary or secondary source of non-Gaussianity without compressing all the information down to a single number. While the bispectrum contains up to $\ell_{\rm max}^2$ independent numbers (due to the triangle condition), the skew spectrum - described with up to $\ell_{\rm max}$ numbers -  offers a useful data compression, while retaining the power to differentiate between different models. A principal component analysis formalises this power through the identification of orthogonal modes. In this section we consider the skew-spectra for various models of primordial non-Gaussianity and for secondary sources of non-Gaussianity. In Figure~\ref{fig:skewspec} we plot the ``{\em off-diagonal}" contribution to the Fisher matrix given by the second term on the right hand side of equation (\ref{eq:covopt}).
This gives an indication of the correlation between different modes in the various models. 
In Figure~\ref{fig:skewspec2} we plot the skew spectra for three primordial models - local, equilateral and orthogonal - as well as for three secondary models - arising from the cross correlation of ISW and lensing (ISW x Lens), thermal Sunyaev-Zeldovich with lensing (SZ x Lens) and residual point sources (PS). As a visual diagnostic it is clear that the skew spectrum allows for the models to be distinguished. However, we are interested here in what information the skew spectrum carries for the individual models. One may expect that the local model requires fewer modes for its description that the others. Having performed the PCA, in Figure~\ref{fig:eigen} we plot the quantity $1/\sqrt{3\sum_{n=1}^i \lambda_n}$ in order to establish the number of modes required for convergence (for $i=\ell_{\rm max}$ this quantity is $\delta f_{NL}$). In this respect one should have in mind the partial-wave decomposition utilised in the Planck non-Gaussianity analysis, whereby approximately $\sim600$ partial waves are required for convergence to be achieved. One may hope that the skew-spectrum gives independent information at each harmonic scale $\ell$, thus ensuring a clearer diagnostic tool to differentiate between different models. From Figure~\ref{fig:eigen} it is apparent that while convergence to within $\mathcal{O} (10\%)$ of the signal to noise is achievable with approximately $400$ modes, the full range of modes is generally required for full convergence. Hence the skew-spectrum $S_\ell$ generally provides up to $\ell_{\rm max}$ independent measures for each model considered.
 One should caution that a visual inspection of this plot does not distinguish between the different models, since the eigenvectors associated with the eigenvalues in each case are likely to differ. To do so would necessitate a joint analysis as described in \textsection\ref{sec:joints}, or in a more robust and efficient manner, to follow the prescription at the end of \textsection\ref{sec:principal} to identify the orthogonal directions and extract the skew spectra for these directions, before quantifying the results in terms of the individual models. Nevertheless, it is instructive to consider the Fisher matrix of the cross-skew-spectrum between models ${\rm X}$ and ${\rm Y}$, i.e. $[{\bf{F}}]^{(\rm X Y)}_{\ell \ell'}$ and perform a principal component analysis to identify the number of modes required for its accurate measurement. For this purpose we consider the cross spectrum between various primary and secondary templates with the equilateral model of non-Gaussianity. In Figure \ref{fig:eigen} we observe that an accurate identification of the overlap between the skew spectra between the equilateral and the local and orthogonal models of primordial non-Gaussianity again requires almost all modes, while the spectral overlap between the equilateral model and the ISW x Lens or SZ x Lens bispectra may be identified with as few as $\sim100$ modes. 

This analysis indicates that, as with the partial wave analysis of Planck data, one may expect the measurement of individual skew-spectra to require measuring $\mathcal{O}(\ell_{\rm max})$ modes. An advantage of the approach adopted here is that the analysis is performed in multipole space. As such features in the data may be identified with specific multipoles.

\begin{figure}
\centering
\large{\textrm{\;\;\;\;Skew-spectra of the eigenmodes}}\par
\vspace{-0.1cm}
\hspace{0.1cm}
{\includegraphics[width=0.8\linewidth]{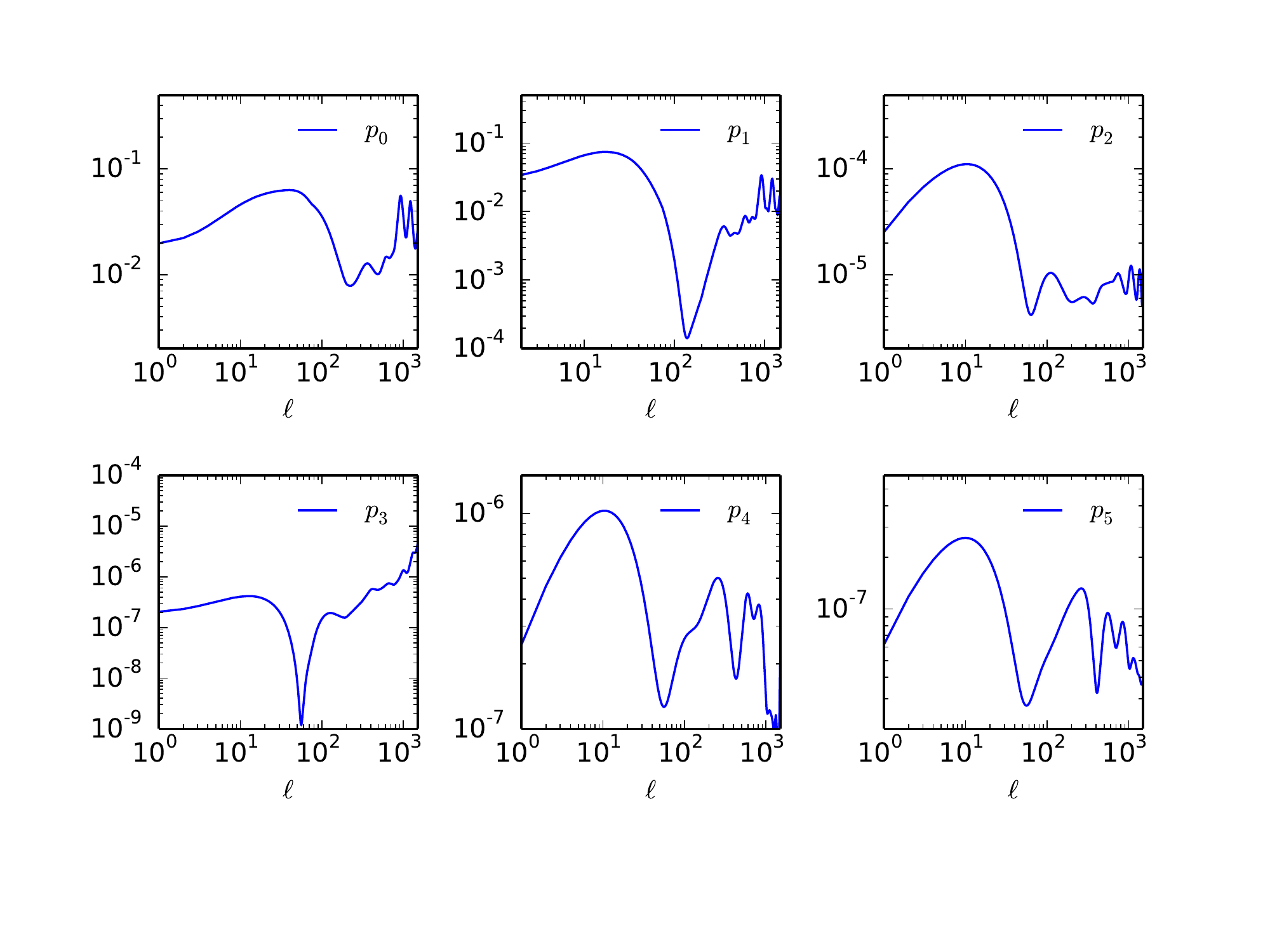}}
\vspace{-1.2cm}
\caption{The skew-spectra of the eigenmodes obtained from a PCA analysis of the six shapes - comprising of three primordial (local, equilateral, orthogonal) and three secondary (ISW-Lens, tSZ-Lens, point source) bispectra.}
\label{fig:skewspec3}
\end{figure}

\subsection{Results for Multiple Skew-Spectra}
A principal component analysis is most useful for the identification of orthogonal directions, ranking them according to those which may be best measured using the data. We consider the case firstly of only primordial models, specifically the local, equilateral and orthogonal models. These three templates are generally regarded as identifying the most distinct shapes that may be constrained. As we show however, the orthogonal directions are given by different combinations of these three shapes. In Table \ref{table:eigenvectors1} we list the eigenvectors corresponding to these eigenvectors, along with the expected error bars (we also list the error bars for each of the templates). We also list the bias induced by each of three secondaries on these primordial models, namely the secondary bispectra ISW x Lens, PS x Lens and SZ x Lens, respectively. We compute also the bias (labelled $\delta_b^{S_j} f_{\rm NL}^{p_i}$) for each of the principal components, $P_i$, due to each of the secondaries, ${\cal S}_j$.
\begin{table}
\centering
	\begin{tabular}{l @{ } c c c c c@{ } r r r r r}
	\hline
	\hline
		\multicolumn{1}{c}{Shape}& \multicolumn{1}{c}{$p_0$} & \multicolumn{1}{c}{$p_1$} & \multicolumn{1}{c}{$p_2$} & \multicolumn{1}{c}{$\delta f_{\rm NL}$} & \multicolumn{1}{c}{$\delta_{b}^{{\cal S}_1} f_{\rm NL}$} & \multicolumn{1}{c}{$\delta_{b}^{{\cal S}_2} f_{\rm NL}$} & \multicolumn{1}{c}{$\delta_{b}^{{\cal S}_3} f_{\rm NL}$}   \\
		\hline
\rowcolor[gray]{0.85}Local &$0.9939$&$0.0981$&	$-0.0503$&$8.14$&$10.2$&$0.46$&$4.88$\\
Equil &$0.0256$& $0.2384$  &  $0.9708$&$76.2$&$2.08$&$1.88$&$17.8$\\
\rowcolor[gray]{0.85}Orthog \,\,&$-0.1072$&$0.9662$&   $-0.2344$&$40.9$&  $-33.1$&$-6.99$&$-71.6$\\
		\hline
		$\sigma_i$&$8.09$&$47.4$&$82.9$&{}&{} \\
\hline
\hline
\end{tabular}
\caption{Eigenvectors of the Fisher matrix chosen from three primordial templates (local, equilateral, orthogonal). The principal directions reassuringly agree well with these templates - especially for the local model. However it is apparent that results quoted for the equilateral and orthogonal models are not independent (though this overlap is to expected). Also listed are error bars for each of the templates and for the principal directions. We also include the bias parameters{, \eqref{eq:bias}}, for each model due to overlap between these primordial models and the ISW x Lens, PSxLens and SZ x Lens secondary bispectra (labelled ${\cal S}_1$ to ${\cal S}_3$, respectively).   \label{table:eigenvectors1}}
\end{table}
\begin{table}
\centering
\begin{tabular}{l c c c c}
         	\hline
                \hline
                Bias & $p_0$& $p_1$& $p_2$ \\
                \hline
		\rowcolor[gray]{0.85}$\delta_b^{{\cal S}_1}f_{\rm NL}^{p_i}$&$10.2$&$-8.86$&$-19.0$\\
		$\delta_b^{{\cal S}_2}f_{\rm NL}^{p_i}$&$0.48$&$-7.40$&$6.53$\\
		\rowcolor[gray]{0.85}$\delta_b^{{\cal S}_3}f_{\rm NL}^{p_i}$&$5.10$&$-75.2$&$64.1$\\
		\hline
		\hline
\end{tabular}
\caption{The bias for each of the principal components
due to the secondaries labelled as ${\cal S}_1$, ${\cal S}_2$, ${\cal S}_3$ that denotes ISW x Lens, PSxLens and SZ x Lens respectively
are shown. }
\end{table}
\begin{table}
\centering
	\begin{tabular}{l @{ } c c c c c@{ } r r r }
	\hline
	\hline
		\multicolumn{1}{c}{Shape}& \multicolumn{1}{c}{$\hat{f}_{\rm NL}$} & \multicolumn{1}{c}{$\delta f_{\rm NL}$} & \multicolumn{1}{c}{$\Delta_{b}^{{\cal S}_1} f_{\rm NL}$} & \multicolumn{1}{c}{$\delta_{b}^{{\cal S}_2} f_{\rm NL}$} & 
\multicolumn{1}{c}{$\delta_{b}^{{\cal S}_3} f_{\rm NL}$}   \\
		\hline
DBI (exact) &$11$&$68.2$&	$ 9.7$&$4.0$&$39.5$\\
\rowcolor[gray]{0.85}DBI (PCA approx)\,\, \,& $-31$  &$68.5$&  $6.7$&$3.6$&$35.8$\\
		\hline
		\hline
	\end{tabular}
\caption{Comparison of the parameters associated with the DBI model computed using the DBI model itself, compared to the approximated form computed from use of the principal components. The values for $\hat{f}_{\rm NL}$ are computed using values reported by the Planck team, but all other results are computed using the simplified Planck-like data described in this section.\label{table:DBIcomparisons} }
\end{table}

As an example of the advantages of using the PCA analysis, we use these results to estimate the error bar for the DBI model using the prescription described in \textsection\ref{sec:principal}, comparing to the error bars computed using the Fisher matrix computed for the DBI model itself. We express the model under consideration in terms of the principal components of the three shapes using equation~\eqref{eq:MbyPCA}. The results using the approximation using the PCA analysis are listed with comparison to the exact results. The estimates for $\hat{f}_{\rm NL}$ are compared to those given in the Planck results paper, \cite{Planck13}, i.e. $\hat{f}_{\rm NL}^{\rm DBI}=11$, with the PCA values computed using those for the templates considered. From Table~\ref{table:DBIcomparisons} we observe that the approximation to the DBI model using the principal components gives a very accurate approximation to the bias parameters (accurate to within $0.05\sigma$) for each of the parameters except for the estimate $\hat{f}_{\rm NL}$ - although we caution, that this value was obtained using central values using the reported Planck results for each of the templates (and for the DBI model itself). The value obtained for the PCA approximation to $\hat{f}_{\rm NL}$ appears to be driven by the degree of overlap between the DBI and equilateral model ($98.3\%$ correlation).

\par
Next we combine the primary and secondary models in order to more accurately characterise the orthogonal directions that may be constrained using the data. In Table \ref{table:eigenvectors2} we again list the corresponding eigenvectors. It is apparent that there is a high degree of overlap between the ISW x Lens and SZ x Lens bispectra as expected. There is also a relatively large overlap between features in the equilateral and orthogonal models. The parameter $\delta f_{\rm NL}$ for the secondaries should be interpreted for the ISW x Lens and the SZ x Lens as representing the inverse of the signal to noise, e.g. for the ISW x Lens model the signal to noise expected at $\ell_{\rm max}\sim1500$ for the Planck satellite is expected to be $\sim 5$. For the residual point source bispectrum $\delta f_{\rm{NL}}$ represents the amplitude $b^{\rm PS}/10^{-29}$.

\begin{table}
\centering
	\begin{tabular}{c @{ } r c c c c c c@{ } r}
	\hline
	\hline
		\multicolumn{1}{c}{Shape}& \multicolumn{1}{c}{${\rm P}_0$} & \multicolumn{1}{c}{${\rm P}_1$} & 
\multicolumn{1}{c}{${\rm P}_2$} & \multicolumn{1}{c}{${\rm P}_3$} & \multicolumn{1}{c}{${\rm P}_4$} & \multicolumn{1}{c}{${\rm P}_5$} & \multicolumn{1}{c}{$\delta f_{\rm{NL}}$}  \\
		\hline
Local &$0.0041$	&$-0.0040$&	 	$0.9942$&	$-0.0210$&	$0.0889$&	$-0.0568$&$8.14$	\\
\rowcolor[gray]{0.85}Equil &$0.0001$&	$0.0001$&		$0.0270$	&	$-0.0048$&	$0.3050$&	$0.9520$&$76.2$	\\
Ortho &$-0.0012$	&$-0.0006$&		$-0.1038$&	$-0.1311$	&	$0.9396$&	$-0.2987$&$40.8$	\\
\rowcolor[gray]{0.85}ISWxLens &$0.6188$&	$-0.7855$&	$-0.0057$&	$-0.0018$&	$-0.0005$&	$0.0003$&$0.20$	\\
SZxLens&$0.7855$&	$0.6188$&	$-0.0008$&	$0.0051$	&	$0.0018$	&$-0.0006$&$0.18$	\\
\rowcolor[gray]{0.85}Point Source\,\,\,&$0.0030$&	$0.0048$&	$-0.0074$&	$-0.9911$&	$-0.1276$&	$0.0361$ &$20.1$\\

		\hline
		$\sigma_i$& $0.16$ & $0.23$ & $8.36$ & $24.0$ &$51.0$&$84.9$&{}\\
		\hline
		\hline
	\end{tabular}
\caption{Eigenvectors of the Fisher matrix for $6$ shapes chosen from three primordial templates (local, equilateral, orthogonal) and three secondary shapes (ISW x Lens,  SZ x Lens and residual point sources).\label{table:eigenvectors2}}
\end{table}
\par
\begin{figure}
\centering
\large{\textrm{\;\;\;\;Eigenvalues corresponding to the principal eigenmodes}}\par
\vspace{0.25cm}
\hspace{0.1cm}
{\includegraphics[width=0.6\linewidth]{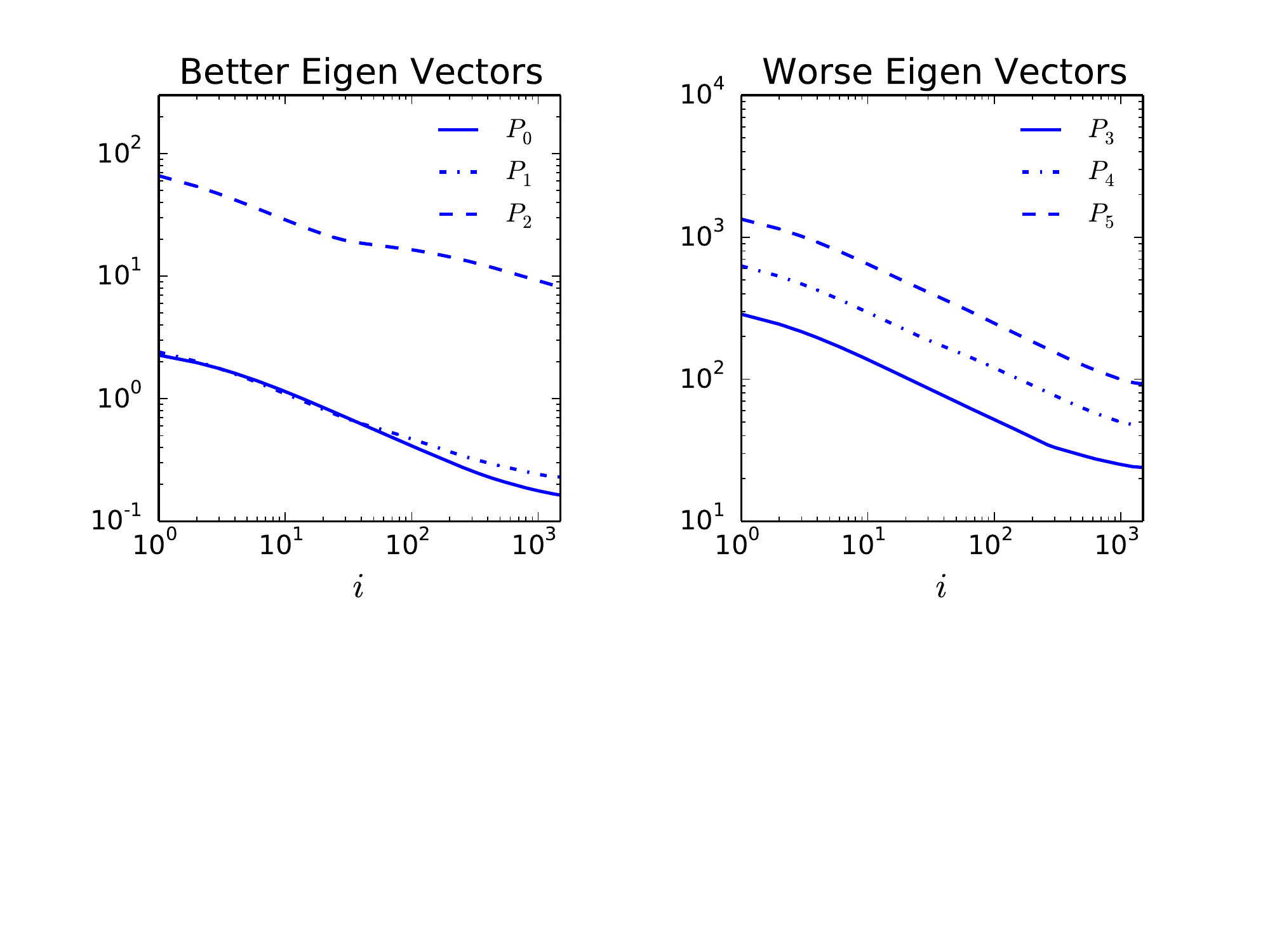}}
\vspace{-3.cm}
\caption{The inverse of the square root of the sum of eigenvalues ($1/\sqrt{\sum_n^i \lambda_n}$) for the two best (left panel) and two worst (right panel) eigenvectors are shown.}
\label{fig:bestworst}
\end{figure}

We now consider the skew-spectra of the principal directions which may be best constrained from the data as listed in Table \ref{table:eigenvectors2}. In Figure~\ref{fig:skewspec3} we plot the associated skew spectra, while in Figure~\ref{fig:eigen} we plot the inverse of the eigenvalues obtained from a decomposition of these spectra. The initial principal component analysis of the six shapes proved useful in identifying the orthogonal directions, while the subsequent decomposition allows for a simpler comparison of data to the skew spectrum of each principal component using the eigenvalues associated with each skew spectrum, $S_\ell^{{\rm P}_i}$. In Figure~\ref{fig:bestworst} we plot the inverse of the square root of the sum of eigenvalues of this subsequent decomposition for the two best and two worst constrained shapes ${\rm{P}}_i$. Again it is apparent that each shape contains up to $\ell_{\rm max}$ independent pieces of information. 
The analysis becomes simpler, since each component is no longer correlated. From the list of eigenvectors, and eigenvalues one may compute the bias simply. For example, the local bispectrum, ${\rm B}_0$, satisfies ${\rm B}_0=\sum_i {\rm W}_{i 0} {\rm P}_i$ and the ISW x Lens model, ${\rm B}_3$, satisfies ${\rm B}_3=\sum_i {\rm W}_{i 3} {\rm P}_i$. Thus the bias due to ISW x Lens is given by $\delta_b f_{\rm NL}=\sum_i {\rm W}_{i 0}{\rm W}_{i 3}\lambda_i^2/\sum _i {\rm W}_{i 0}^2\lambda_i^2$.
Since the primordial directions are not impacted heavily by the presence of secondaries (compare the eigenvectors of Table~\ref{table:eigenvectors1} and ${\rm P}_2$, ${\rm P}_4$ and ${\rm P}_5$ of Table~\ref{table:eigenvectors2}) one may compute the approximate form of a primordial model using the the PCA analysis of only the three primordial templates using Table~\ref{table:eigenvectors1} in order to write the model as in the form ${\rm M}=\sum_{i=0}^2 a_i {\rm B}_i$ and compute the expected bias by reference to Table~\ref{table:eigenvectors2}. 

\section{Discussion and Conclusions}
\label{sec:conclu}
A formalism has been developed in this paper to exploit the use of a principal component analysis to assist in the study of both primary and secondary sources of non-Gaussianity. The PCA analysis is used in conjunction with the skew spectrum statistic with which the bispectrum is represented with a pseudo-${\cal C}_\ell$ representation, and so in terms of $\ell_{\rm max}$ numbers. The advantage of the skew spectrum, $S_\ell$, is that, while the statistic remains an optimal method with which to measure non-Gaussianity, the information is not compressed to a single number, $f_{\rm NL}$. In addition one may associate features with a harmonic scale, $\ell$. However, there remains the possibility that perhaps a sufficient statistic with which to accurately identify a source of non-Gaussianity may require far fewer numbers. PCA formalises into a clearly defined problem. 

We apply the principal component analysis to skew spectra associated with individual bispectra (both primary and secondary), to the skew spectra associated with the three Minkowski functionals and to the joint analysis of multiple bispectra. The PCA may be described as a method with which to identify the uncorrelated components in each of the cases. In this paper we apply our analysis to the case of Planck-like data, with realistic estimates of the beam and noise. 

First we considered the case of individual skew-spectra associated with the three standard primordial templates (local, equilateral and orthogonal) and three secondary models - namely those due to the cross correlation of lensing with the integrated Sachs Wolfe effect, due to the cross correlation of lensing with the thermal Sunyaev-Zeldovich and due to residual point sources, respectively. Having computed the eigenvalues associated with the PCA decomposition, we found that the skew-spectrum generally provides up to $\ell_{\rm max}$ independent pieces of information. One should compare to the modal analysis employed by the Planck team which requires a similar number of modes for convergence. A particular advantage of the approach adopted here is that features may be associated with specific harmonic scales.

Next, the techniques are applied to the three skew spectra associated with the Minkowski functionals. Though a suboptimal estimator, these topological estimators allow for a useful cross check of the non-Gaussian statistics. 

Next, we consider the joint estimation of primary and secondary bispectra - again analysing the three standard primordial templates, along with the ISW x Lens, SZ x Lens and the residual point source bispectra. Considering firstly the three primordial models in isolation, we identify the orthogonal directions. We utilise the results of the earlier sections to make estimates of the bias due to the secondaries.
We describe how the PCA may be used to relate the results reported using Planck data on the standard primordial templates to infer the constraints on another primordial model without requiring a full analysis to be carried out in that case (all that is required is the calculation of the Fisher matrix of that model along with a measurement of the correlation between this model and the standard templates - this correlation may be approximated using a simple primordial measure). We have applied this approximation to the case of the DBI bispectrum to obtain constraints on this model as well estimates of the bias induced by three secondary models. Having identified the orthogonal directions, we next considered the skew spectra associated with each of the principal components. Performing a principal component decomposition on these skew spectra allows one again to identify the (up to) $\ell_{\rm max}$ independent pieces of information that may be measured.

The formalism described allows for an efficient method with which to analyse the skew-spectra associated with a joint analysis of primary and secondary bispectra. The application to active sources of non-Gaussianity such as cosmic strings \cite{Hind09,Regan09} is straightforward. The use of a principal component analysis may prove useful in identifying any underlying source of non-Gaussianity. This may be particularly useful in the context of recent work on the possibility of using the skew spectrum of the bispectrum associated with the ISW x Lens signal to probe for modified theories of gravity \cite{Munshi14}.  

We have ignored iscurvature perturbations in this paper but the
PCA analysis of joint estimates using adiabatic and isocurvature bispectrum can be performed in a similar manner to the work presented here.
The PCA results presented here can also be extended to higher-order i.e. to the kurt-spectra. 
It will also be possible to include skew-spectra associated with mixed bispectra involving polarization and temperature anisotropies.

\section*{Acknowledgements}
DM and DR acknowledge support from the Science and Technology Facilities Council
(grant numbers ST/L000652/1 and ST/I000976/1 respectively). DR is also supported by funding from the European Research Council under the European Union's Seventh Framework Programme (FP/2007-2013) / ERC Grant Agreement No. [308082]. Much of the numerical work presented in this paper was obtained using the COSMOS supercomputer, which is funded by STFC, HEFCE and SGI.
It's pleasure for DM to thank Bin Hu, Alan Heavens and Peter Coles for related collaborations and useful discussions. DR wishes also to thank Antony Lewis for helpful comments on a draft of this work.
\label{acknow}

\bibliography{potenBIB.bbl}
\end{document}